\documentclass[traditabstract]{aa}
\usepackage{graphicx}
\usepackage{subfloat}
\usepackage{graphicx}
\usepackage{caption}
\usepackage{subcaption}
\usepackage{amsmath}
\usepackage{natbib}
\usepackage{color}
\usepackage[english]{babel}
\bibpunct{(}{)}{;}{a}{}{,} 
\newcommand{\Msun}{M\ensuremath{_\odot}\,}
\begin{document}
\title{
Phylogeny of the Milky Way's inner disk and bulge populations: Implications
for gas accretion, (the lack of) inside-out thick disk formation, and quenching
} 

\titlerunning{Phylogeny of the Milky Way's inner disk and bulge populations}

\author{Misha Haywood\inst{1} 
\and Paola Di Matteo\inst{1}
\and Matthew Lehnert\inst{2}
\and Owain Snaith\inst{3}
\and Francesca Fragkoudi\inst{1}
\and Sergey Khoperskov\inst{1}
}

\institute{GEPI, Observatoire de Paris, PSL Research University, CNRS, Sorbonne Paris Cit\'e, 5 place Jules Janssen, 92190 Meudon, France\\
\email{Misha.Haywood@obspm.fr}
\and
Sorbonne Universit\'{e}, CNRS, Institut d'Astrophysique de Paris, 98 bis bd Arago, 75014 Paris, France
\and 
School of Physics, Korea Institute for Advanced Study, 85 Hoegiro, 
Dongdaemun-gu, Seoul 02455, Republic of Korea  
}

\abstract{
We show that the bulge and the disk of the Milky Way (MW)
at R$\lesssim$7~kpc are well described
by a unique chemical evolution and a two phase star-formation history
(SFH). We argue that the populations within this inner disk, not the
entire disk, are the same, and that the outer Lindblad resonance  (OLR) 
of the bar plays a key role in explaining
this uniformity.  In our model of a two phase star formation history, the
metallicity,  [$\alpha$/Fe] and  [$\alpha$/H] distributions, and age-metallicity relation 
are all compatible with the observations of both the inner disk and bulge.
The dip at [Fe/H]$\sim$0 dex seen in the metallicity distributions of the
bulge and inner disk reflects the quenching episode in the SFH of the
inner MW at age $\sim$8 Gyr, and the common evolution
of the bulge and inner disk stars.  Our results for the inner region,
R$\lesssim$7 kpc, of the MW, are consistent with a rapid build-up of a large
fraction of its total baryonic mass within a few Gyr. We show that at z$\le$1.5, when
the MW was starting to quench, transitioning between the end of the $\alpha$-enhanced thick
disk formation to the start of the thin disk, and yet was still gas rich,
the gas accretion rate could not have been significant.  The 
[$\alpha$/Fe] abundance ratio before and after this quenching phase would be
different, which is not observed. The decrease in the accretion rate and gas fraction
at z$\le$2 was necessary to stabilize the disk allowing the transition
from thick to thin disks, and for beginning the secular phase of the
MW's evolution. This possibly permitted a stellar bar to develop which we
hypothesize is responsible for quenching the star formation. 
The present analysis suggests that Milky Way history, and in particular at the transition from the thick 
to the thin disk -- the epoch of the quenching -- must have been driven by a decrease of the star formation 
efficiency.
We argue that the decline in the intensity
of gas accretion, the formation of the bar, and the quenching of the SFR at
the same epoch may be causally connected thus explaining their temporal
coincidence. Assuming that about 20\% of the gas reservoir in which metals are diluted
is molecular, we show that our model is well positioned on the Schmidt-Kennicutt relation at all times.
}


\keywords{Galaxy: abundances --- Galaxy: disk --- Galaxy: evolution --- galaxies: evolution}
\maketitle


\section{Introduction}

The stellar-mass density distribution of galaxies shows that a majority of stars in the local 
universe are in galaxies that have a mass similar to that of the MW \citep[e.g.][]{papovich2015,bell2017}. Recent studies have 
robustly established that these galaxies form half their mass before z$\sim$1, or about 8 Gyr ago 
\citep{muzzin2013, vandokkum2013, patel2013, papovich2015}. The MW closely follows this general 
behavior \citep{snaith2014} and for our Galaxy, the thick disk alone can explain the early (within the first 3-4 Gyr) 
steep mass growth observed in Milky Way-type galaxies. As emphasized recently by \cite{bell2017}, 
galaxies with stellar mass similar to that of the MW come with a wide variety of morphologies in the local universe, and it is not clear 
which stellar populations are responsible for this mass growth, in particular because both classical bulges 
and thick disks are liable to form at these epochs.

In the MW, while evidence for a pseudo-bulge -- or a bulge being a bar formed from dynamical 
instabilities in the disk -- is now overwhelming, it is still not entirely clear how the disk 
contributed to its formation, because of discrepancies in the properties of the bulge and the disk, 
and in particular of their chemical properties (see, for a variety of point of views, e.g. 
McWilliam 2016, Di Matteo 2016, Zoccali 2016, Bensby et al. 2017, Haywood et al. 2016, 
Johnson et al. 2013), and the known difficulty to fit a single chemical evolution model to both the 
bulge and disk \citep[e.g][]{ballero2007,tsujimoto2012}. Hence, when estimating the mass growth of 
the MW, it may not be entirely clear how the bulge should be counted, because it depends on what 
population it is exactly. The concept of stellar population has come in various flavors in the recent 
literature,  with 'geometrically' \citep[e.g.][]{martig2016} or 'chemically' \citep{bovy2012b} defined 
stellar population, which, for the bulge particularly, may be confusing.

A practical definition 
of a stellar population is a group of stars that presents some unicities in either their spatial, kinematic, 
chemical properties and age, or several of these properties (see e.g. Haywood et al. 2013 on how these properties 
are related). 
We are looking for a definition that goes beyond the mere identification of stars however, to understand the role
this population has played in the formation and evolution of the Galaxy. 
Starting from observables,  two steps can be seen as necessary to characterize a population according 
to this definition: (a) find a criterion unambiguously linked to evolution according to which we can 
identify stars; (b) relate these stars to a known phase of galactic evolution. For the identification of 
a population to be robust, the criteria used in (a) must be reliable. 
Some time after the formation of a population however, spatial and kinematics properties become second order 
signatures  of a fossil record of a population \citep{freeman2002}, because various dynamical processes 
(e.g bar formation, satellite interactions) are able to reshape distributions, blurring the initial spatial and kinematic signatures. 
In the recent years, chemical abundances have become the genetic markers of stellar populations to astronomers 
\citep{freeman2002, fuhrmann1998, reddy2003, reddy2006, haywood2008, nissen2010, haywood2013}, and much work is actually 
being done to both acquire large swaths of new spectroscopic information and to relate this information to Galactic 
evolution (APOGEE I\&II, WEAVE, 4MOST, MOONS), through proposed techniques such as chemical labelling or tagging \citep{bland2004}.
Chemical tagging aims at identifying the molecular cloud from which a star originates, thereby identifying the most 
elementary unit of Galactic evolution. We wish on the contrary to relate the chemical identity of a star to a more ``holistic'' information.
We have shown in \cite{snaith2014, snaith2015} that the alpha abundances of a population and their evolution with 
time are directly related to its star formation history, or in-situ mass growth. If each step of the mass growth of 
our Galaxy can be related to a corresponding stellar population, through the process described in (a) and (b), then our definition of stellar populations is meaningful and will help us to 'reconstruct' the processes out of which our MW formed. 

Having clarified the general definition of a stellar population we intend to use, it must be said that, in trying to reconstruct the mass growth history of our Galaxy, we are led to question the parenthood and relationships between the groups of stars that we recognise as populations, in a process akin to phylogeny in biology. 
This is particularly acute in the MW for the thick disk and bulge, and a now large literature illustrates the question: are the thick disk and bulge the same population ? \citep[][to name just the first papers on the subject]{melendez2008, alvesbrito2010, bensby2010, ryde2010}.
In trying to answer this question in the following pages, we rely on the same definition of the thick disk as the one given in Haywood et al. (2013), where we showed that a chemically-based definition for the thick disk (using alpha element abundances) is the most directly linked to its evolution, because of the tight relation that links alpha abundances to ages, as found in that article.
Using this definition, the thick disk considered in this work is unambiguous and concerns stars that are essentially confined to R$<$10kpc, 
are old ($>$9~Gyr) and are distributed in a thicker disk than the thin disk. We refer the reader to our discussion in Haywood et al. (2013) about the ambiguities of the thick and thin disks nomenclature.

Our article builds on these concepts by showing that there is strong evidence that the bulge and the {\it inner} disk (which we define as the thick {\it and} the thin disk inside R$\lesssim$ 7~kpc) are essentially the same population, and that their chemical evolution can be described by a single and simple model. 

The outline of our paper is the following.
In the next section, we present our chemical evolution model, and the scenario that goes with it and 
that was set up in a series of recent papers by our group.
The model is confronted to the inner disk constraints in 
Section \ref{sec:disk}, and to the bulge observations in Section \ref{sec:bulge}. In Section \ref{sec:thesame} 
we discuss arguments that have been invoked to claim that the inner disk and the bulge are the same population, 
while Section \ref{sec:accretion} presents a discussion about the accretion history of the inner MW. Our conclusions are given in the last section.

\section{Scenario and model}\label{sec:scenario}

For reasons that are discussed in Section \ref{sec:accretion}, we want to test if the inner MW
can be represented by a scenario where most gas accretion has occured early in the evolution of our Galaxy -- or before significant 
star formation started. The model that best approximates this situation is the closed-box model.
Our aim in the present study is to understand how far this simple modelling may be a valid representation of the data to R$\sim$6-7 kpc.
It is therefore applied to the whole inner disk-bulge stellar system.
We first describe the context and limits in which we apply this model, then the ingredients and characteristics of the model itself.

\subsection{Landmarks in the Milky Way history and a scenario}

Our study is developed in a context framed by a number of different articles: \cite{haywood2013,lehnert2014, dimatteo2014, snaith2014,snaith2015, halle2015, haywood2016a, haywood2016b, dimatteo2016} but also numerous articles from the literature cited below. We now summarize this context.

(1) In \cite{haywood2013}, we showed that the thick disk formed starting from about 13 Gyr ago to about 9 Gyr, during the most intense phase of star formation in the MW \citep{snaith2015}, in a star burst mode \citep{lehnert2014},  as can be traced back from the age-alpha relation.
The chemical homogeneity of this population \citep{haywood2013, haywood2015} at all ages implies a high level of turbulence in the gas that formed this population. 

(2) Fitting the observed age-alpha relation of the inner disk with our GCE model \citep{snaith2014, snaith2015} we were able to infer its SFH showing two distinct phases. 
The thick disk was formed in the first phase (13-9 Gyr), and represents half the stellar mass of the disk. The thin disk is formed
in the second phase, from 7 Gyr to the present day. During the early thick disk phase (age $>$11 Gyr), we found that the SFR density was sufficiently high to generate possible outflows \citep{lehnert2014} that may have contributed to pollute the outer ($>$10kpc) disk \citep{haywood2013}.

(3) \cite{haywood2013} proposed that the disk is better defined by the dychotomy of its radial structure (inner/outer disk) than vertically (thin/thick disk). This was confirmed afterwards by in-situ APOGEE data (Hayden et al. 2015). 
We argued in \cite{halle2015} that, at the formation of the bar, which in galaxies of the mass of the MW, typically occurs at z$\sim$1 \citep{sheth2008, melvin2014}, the OLR established a barrier that essentially isolated the inner disk from the outer disk (apart from blurring effects). As shown in \cite{dimatteo2014}, only stars that are within the OLR will participate to the bar and boxy/peanut bulge. Fig. A.1 of this article illustrates with a simulation from \cite{halle2013}  that stars that are beyond the OLR will stay in the outer parts and will not participate to the bar. In the MW, the exact position of the OLR is debated: slightly inside the solar orbit \citep{dehnen2000}, or further away \citep[e.g.][who, after the work of \cite{portail2017}, proposed that it is situated at 10.5 kpc]{perez2017}.
It is however expected that the OLR has moved outward as a consequence of the slowdown of the bar, hence it may have formed within the solar orbit and 
moved further out. 
In \cite{halle2015}, we argued that it has maintained the inner disk essentially separated from the outer 
disk, which developed two separated chemical evolutions. Hence, the OLR delineates two regions of the disk that will mix only a limited amount of their stars. 
The possible effect of the OLR is therefore a fundamental feature of our analysis because it justifies that the bulge and the {\it inner disk} must be the same population, not the bulge and the disk as a whole.

(4) We found in \cite{haywood2016a} that within R$<$6-7~kpc, a major cessation (or 'quenching') of star formation occured at the transition from the thick to thin disks, between 9 to 7 Gyr. This quenching however did not occur because of gas exhaustion, since in this case stars formed after the quenching would show chemical discontinuity with stars formed before the quenching. This is not observed: \cite{haywood2013}
have shown that there is a chemical continuity between the thick and thin inner disks. In \cite{haywood2016a}, we argued that the formation of the bar may have been 
responsible for quenching the star formation. We emphasize that it is the quenching of the star formation rate which allowed the transition from the thick to the thin disk.
In \cite{khoperskov2017}, we have shown that the formation of a bar is indeed able to reduce significantly ($\sim$ a factor 10) and in a relatively short amount of time ($\sim$ 1-2 Gyr)  the SFR of a disc galaxy, in a way at least qualitatively compatible with the MW.

(5) The bulge of the MW is mainly a pseudo-bulge. Its kinematical properties can be explained if it is (mainly) a bar grown from dynamical instabilities in the disk, with a negligible or nonexistent classical bulge \citep{shen2010, kunder2012, ness2013a, ness2013b, dimatteo2014}. Having essentially a disk origin, it must be made of thick and thin disks stars that originate from inside the OLR (Di Matteo et al., 2014, 2015, Di Matteo 2016).
The CMD of the bulge as observed by the HST shows a very tight turn-off \citep{clarkson2008}, suggesting a mono-age, old population. However, it has been shown in \cite{haywood2016b} that the SWEEPS field turn-off can only be compatible with the observed metallicity distribution function (hereafter MDF) of the bulge if there is a correlation between age and metallicity, implying that essentially all super-solar metallicity stars in the bulge must be younger than about 8 Gyr \citep[see also][]{bensby2013, bensby2017, schultheis2017}.

The above results have two general implications. The first one is that if the bulge is a pseudo-bulge formed from the disk inside the OLR, the two must show the same chemical characteristics.
As already said, while the problem of the identification of the bulge and the thick disk has been claimed and investigated several times, the question at stake here is the identification of the inner disk, including both the thin and thick disks, to the bulge.
Second, according to result (2) above, the thick disk represents about half the stellar mass in the inner disk, and this implies that large amounts of gas were available for star formation early (age$>$10 Gyr) in the disk. 
As we argue in section \ref{sec:noinsideout}, this implies significant difference from the assumptions of the inside-out paradigm, and opens the way to use a closed-box model, which we now introduce, as a possible first approximation of this situation.

\subsection{The model}\label{sec:model}

The closed-box model we use -- or more precisely the simple closed-box model, which is the closed-box model
with no radial dependence of the SFH -- has been described in \cite{snaith2015} and assumes (1) a constant IMF (2) an always well mixed ISM (3) no inflow or outflows of gas (4) initially pristine gas with no metals, but no instantaneous recycling approximation, that means we derive the full information provided by detailed yields and metallicity-dependent stellar lifetimes. 

There are three fundamental differences between our model and standard closed-box models:

(1) Closed-box models, when accompanied with a Schmidt-Kennicutt type of star formation law, lead to very fast exhaustion of gas and unrealistic age distributions for a disk (see for instance \cite{fraternali2012}).
In \cite{snaith2014,snaith2015}, we derived the SFH of the MW inner disk by fitting the model to the age-[Si/Fe] relation, but without assuming any Schmidt-Kennicutt law.  As commented in \cite{haywood2015}, because we argued that the age-[Si/Fe] derived from local stars is valid for the whole inner disk, the deduced SFH is also expected to hold for the whole Galaxy inside the OLR. As mentioned previously, the derived SFH presents two distinct phases of star formation, corresponding to the thick and thin inner disks.
Figure \ref{fig:modeldist} shows the various model distributions: the MDF (top plot), the age-metallicity distribution (middle plot), the SFH (bottom plot).
Two models are shown. The thick curves represent the fiducial model obtained in Snaith et al. (2015), while the thin curves represent distributions obtained in the case of a
schematic SFH where we have removed the small scale variations of the best fit model. In the rest of the article, these two models will be referred to as the fiducial and the smoothed models, respectively.

(2) While classically the closed-box model (hereafter CBM) has been designed for and compared to the solar vicinity, its application is considered here for the entire bulge and the inner disk, with the aim to test if they can be described as a unique system.
The condition of the validity of the CBM to describe the solar vicinity is provided in another article (Haywood et al. 2017, in prep.). 
The ``closed-box'' is considered here as interesting because it approximates a system strongly dominated by gas at early times, while the fact that it is closed/open has no bearing in this case. Section \ref{sec:accretion} further discusses how the approximation of the chemical evolution provided by the CBM fits in the context of cosmological gas accretion and Schmidt-Kennicutt law.

(3) The MDF of a closed-box model is known as a bell-shaped distribution, but this is obtained under the approximation that the recycling of gas ejected by stars is instantaneous, and in this case the shape is independent of the SFH \citep[e.g.][]{pagel2009}. When the recycling of chemical elements from stars is properly modeled, as it has been done in \cite{snaith2015}, it can however produce a very different MDF, which depends to first order on the shape of the SFH. Fig. \ref{fig:modeldist} shows the MDF,  age-metallicity relation, and corresponding star formation history  
assuming a CBM obtained in \cite{snaith2015} (thick curve). The model only constraint was to best fit the age-[Si/Fe] relation of solar vicinity stars, which, as we argued in \cite{haywood2015}, should be representative of the whole inner disk.
The MDF shows two distinct peaks at [Fe/H]=+0.3 and -0.2 dex, separated by a dip at [Fe/H]=+0 dex (or [Fe/H]=+0.1 dex for the smoothed SFH), which corresponds to the epoch when the MW, in a similar way as other galaxies, quenched its star formation activity according to \cite{haywood2016a}, as can be seen by the shaded areas in the three plots. 
The peak at [Fe/H]=-0.2 dex is the signpost of the thick disk formation and the peak at [Fe/H]=+0.3 dex corresponds to the stars that formed during the (inner) thin disk phase (age $<$ 7.5 Gyr). 
The smoothed model shows that the third
peak of the MDF at [Fe/H]$\sim$-0.8 dex is due to the small peak in the SFH at age$\sim$12 Gyr, which is within the uncertainties of the SFH
and cannot be taken as significant.

We note that the present model is entirely constrained by fitting the inner disk sequence of the solar vicinity age-[Si/Fe] relation \citep{snaith2015}, and that it is not tuned in any aspect to fit the in-situ inner disk and bulge data. We do not introduce any additional parameter such as a radially dependent star formation timescale (as a consequence of a radially dependent accretion timescale, e.g. \cite{minchev2014, kubryk2015a}) or radial migration, which are essentially unconstrained, and therefore are $de facto$ free parameters. 
We also emphasize that the model presented here applies to the inner disk only: the outer disk and solar vicinity have different age-[$\alpha$/Fe], age-[Fe/H] relations 
and MDFs from the inner (thick+thin) disks and need separate chemical evolution modelling (see \cite{snaith2015} for the modelling of the outer thin disk 
age-[$\alpha$/Fe] relation). A scenario for the chemical evolution of the solar vicinity is presented in Haywood et al. (2018, in preparation).

\subsection{Ingredients}

The ingredients of our model are described and discussed at length in \cite{snaith2015}. 
The standard theoretical yields are from \cite{iwamoto1999},  \cite{nomoto2006}, \cite{karakas2010}, the IMF from 
\cite{kroupa2001}, and the stellar mass--lifetime relation dependent on the metallicity and is taken from \cite{raiteri1996}, and 
the time delay function is taken from \cite{kawata2003}. As already said, no instantaneous recycling approximation has been assumed, 
see \cite{snaith2015} for details.


\section{ Results: Chemical evolution of the inner disk}\label{sec:disk}

\begin{figure}
\includegraphics[trim=10 10 50 80,clip,width=9.cm]{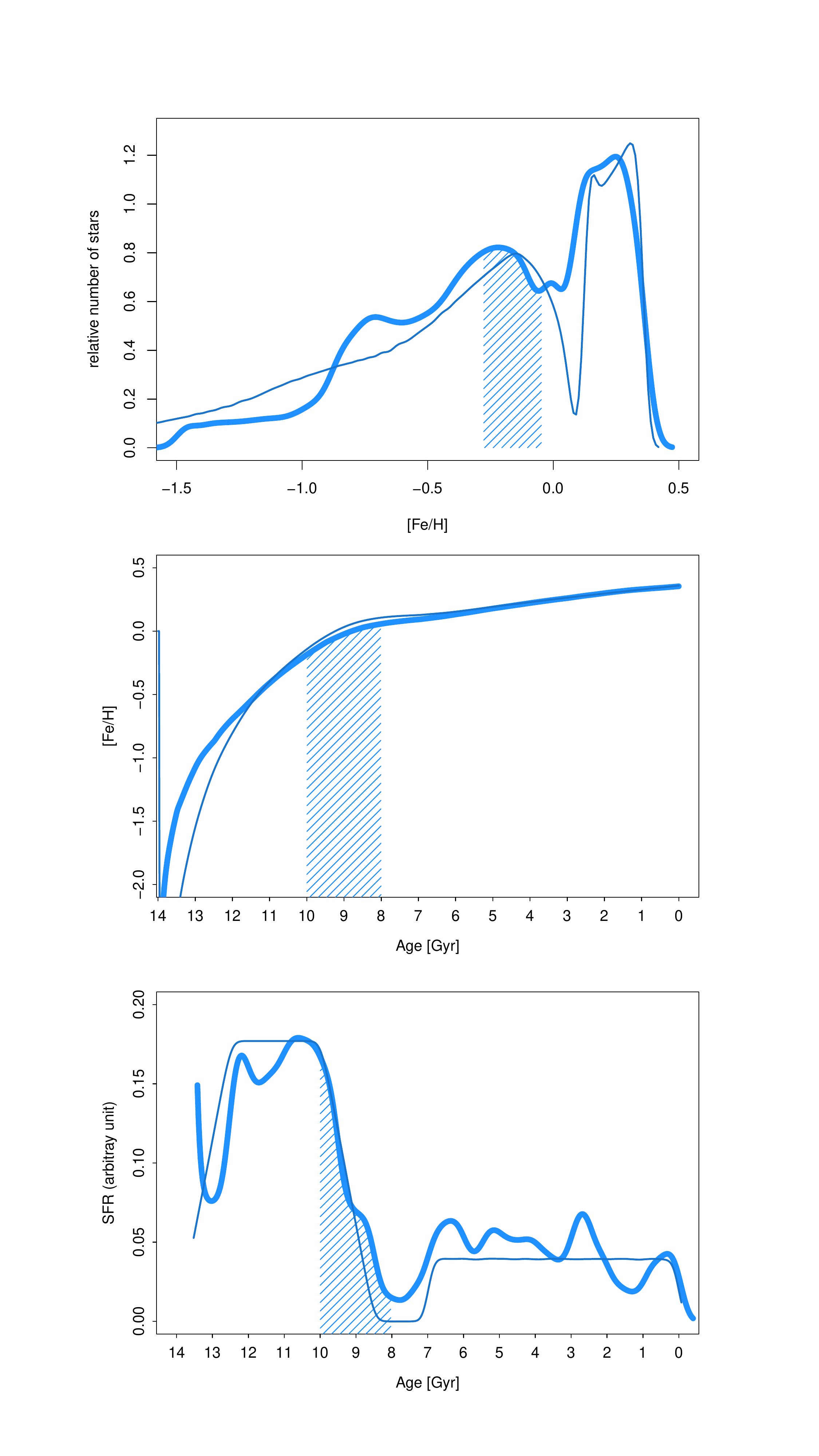}
\caption{The distributions of our inner disk model. Bottom panel: the star formation history; 
Middle: the age-metallicity relation. Top: the metallicity distribution. 
The thick curve corresponds to the best fit model derived in \cite{snaith2015}, or 'fiducial model', while the thin curve
represents the same model, but where the small scale variations in the SFH have been smoothed out.
It shows in particular that the third peak (at [Fe/H]$\sim$-0.8 dex) in the \cite{snaith2015} is
due to the small peak in the SFH at age $\sim$ 12 Gyr, and is probably not significant.
The blue area in each plot
emphasizes the quenching epoch of the SFR in the MDF and age-metallicity relation.
}
\label{fig:modeldist}
\end{figure}


\subsection{Data and the MDF of the inner disk from APOGEE}

The Apache Point Observatory Galactic Evolution Experiment \citep[APOGEE;][]{majewski2015} is a near-infrared, high-resolution ($R\approx 22,500$) spectroscopic 
survey of stars in the MW, included as part of the third and fourth Sloan Digital Sky Survey. In the present study, we use the 
data release 13 of the APOGEE survey. 
Starting from a sample of $\sim$ 104000 stars in APOGEE with distances (and $\sim$ 15000 within R$<$ 6kpc) from \cite{wang2016}, 
we select APOGEE objects as recommended in the DR13 documentation, discarding objects with Teff$>$5250K
or [Fe/H]$<$-1 dex or outside the interval 1.0$<$log g$<$3.8. 
Within these limits, it has been shown by Hayden et al. (2015) that the sampling in directions, magnitudes and colours of APOGEE 
does not introduce any significant bias in the metallicity distribution function of the survey. Giants however probably bias the 
underlying age distribution against the oldest objects, or the most alpha-rich stars, see Bovy et al. (2014). Hence, we 
keep in mind that the survey possibly underestimates the relative number of old stars compared to younger ones, or the number of stars 
in the thick disk.

\begin{figure*}
\includegraphics[width=18cm]{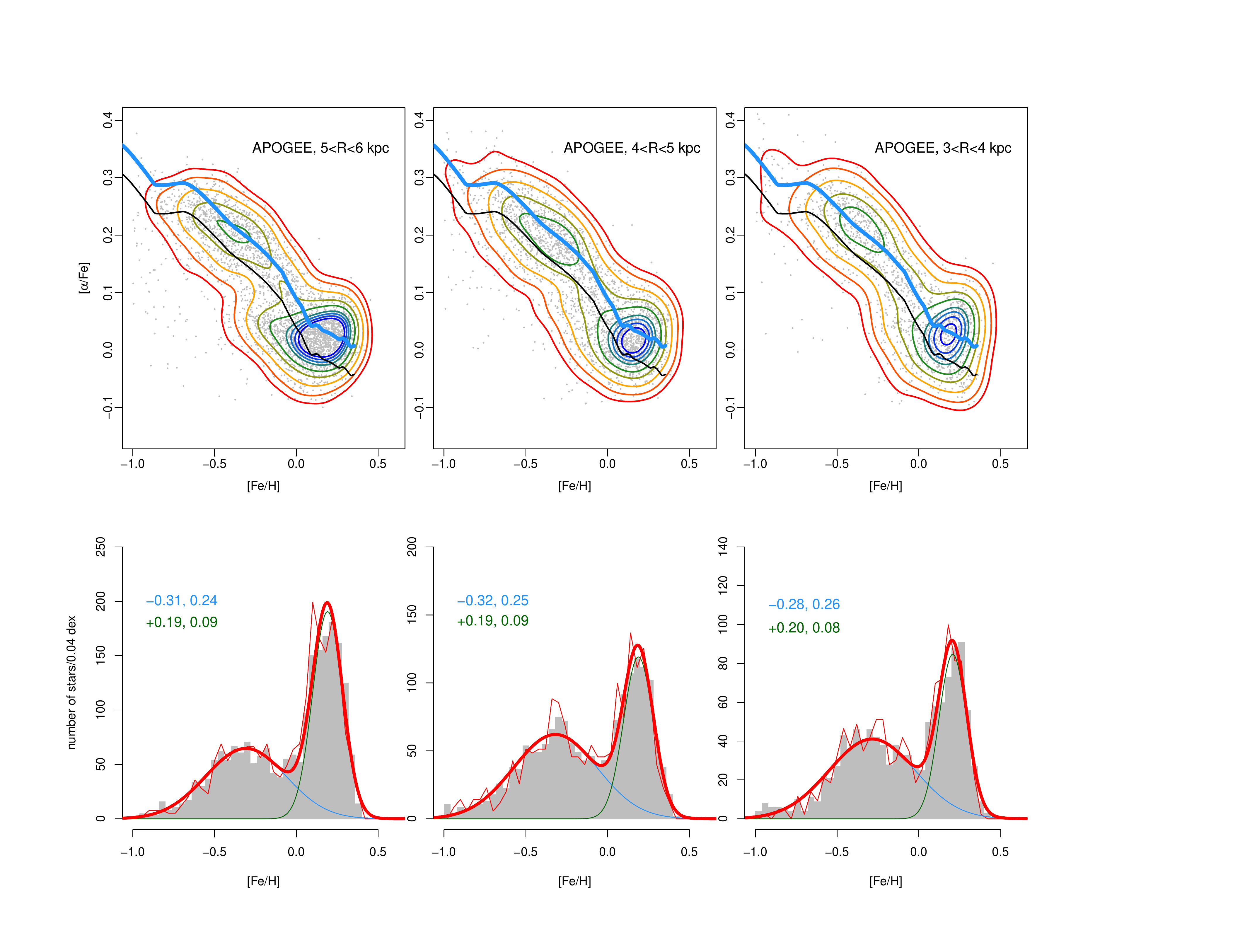}
\caption{Top: the [$\alpha$/Fe]-[Fe/H] distribution of stars from APOGEE in different distance bins as indicated on each plot.
Our standard model is shown in blue, the same model lowered by -0.05 dex in black. Stars that compose our MDFs are 
selected above this line.
Bottom: Gray histograms are MDFs of stars selected above the black line. The thick red curves are the sum 
of the two components (blue and green) in the finite mixture decomposition applied to the gray histograms.
The mean metallicity and dispersion of each component is given on each plot. 
The red thin curves are the histograms obtained by selecting stars in the APOGEE catalogue with a signal to noise ratio above 150 (instead of 50 for the gray histogram).
}
\label{fig:mdfinnerdisk}
\end{figure*}

Fig. \ref{fig:mdfinnerdisk} shows the [Fe/H]-[$\alpha$/Fe] distributions of the APOGEE data
from 3 to 6 kpc from the Galactic center in three different distance intervals, 
assuming the Sun is at 8 kpc from the Galactic center.
[$\alpha$/Fe] is defined by the mean of Mg and Si abundance ratios.
A finite mixture density estimation in this plane (using the Mclust R package) indicates two main components 
around the following mean metallicities and alpha abundances in the three distance intervals, 5-6, 4-5, 3-4 kpc: 
[Fe/H]=-0.40, -0.41, -0.35, [$\alpha$/Fe]=0.20, 0.19, 0.20, for the metal-poor component and [Fe/H]=0.16, 0.17, 0.12, and 
[$\alpha$/Fe]=0.018, 0.017, 0.028, for the metal-rich. 
These values are near the centroids of the two components seen in the density contour plots, Fig. \ref{fig:mdfinnerdisk}.
The first two distance bins (5-6/4-5 kpc) show an additional 
components at [Fe/H]=-0.11, -0.06 dex  and [$\alpha$/Fe]=0.05, 0.05 dex representing respectively 24 and 18\% of the sample, 
which we interpret as contamination by solar vicinity/outer disk stars.
This third component is not found in the third bin (3-4 kpc).
In order to limit the contamination by this last component, we selected stars as in \cite{haywood2016a}, keeping 
objects above our standard model lowered by 0.05 dex (black curve). 
The resulting MDFs are shown on the lower plots as gray histograms. 
This selection limits, but cannot completely eliminate solar vicinity/outer disk stars contamination, particularly 
in the first distance bin, between 5 and 6 kpc. 
We have chosen not to use the  estimated  membership probabilities of individual stars given by Mclust 
because the third component, dominated by outer disk objects, also comprises a number of stars which are clearly on the inner disk sequence. 
Removing them would artificially enhance the bimodality of the distributions.

We also applied the finite mixture density estimation to one-dimensional metallicity histograms (gray histograms, bottom plots of Fig. \ref{fig:mdfinnerdisk}) obtained after the selection of our stars, searching for two and three components.
The bayesian information criterion (hereafter BIC) defined by BIC = -2 ln(L) + k ln(n), (where L is the likelihood of the model, 
n is the number of bins and k the number of parameters to assess the goodness of the fit) shows that in all three distance bins, the data  
is fitted equally well with two or three components. 
For the first and second distance bins, it is either the main metal-rich or the metal-poor peaks which are 
sub-divided in two components, and the BIC information is inconclusive, slightly favoring 3 components 
in the first distance bin, and 2 in the second one. 
In the third distance bin, the 3rd component is only 2\% and has a mean metallicity of
-0.91. 
Hence, the two-components fit offers a more uniform solution, with the mean metallicities of the metal-poor ([Fe/H]=-0.30$\pm$0.02 dex) 
and the metal-rich ([Fe/H]=0.19$\pm$0.01 dex) components very similar in all distance bins 
and dispersions which are about +0.25 dex for the metal-poor component and 0.09 dex for the metal-rich. 
These parameters are very similar to the maximum of the density contours shown in the [Fe/H]-[$\alpha$/Fe] distribution, 
and to the parameters of the two main components found in the 2D analysis above, with only a limited increase in the mean metallicity
of the metal-poor component by about +0.05-0.1 dex.
Hence, our selection procedure keeps intact the main characteristics of the 2D distributions.
In all distance intervals, the transition between the metal-poor and metal-rich peaks is found at solar metallicity. 
This is very similar to the result obtained by several groups on the bulge, where 
two dominant peaks are most often found, see for instance \cite{zoccali2017}.
In all plots, stars have been selected with a minimum signal to noise ratio of 50. 
We checked that changing the limit does not affect significantly the 
shape of the MDF, apart from increasing the Poisson noise in the histogram. The effect of
setting the signal-to-noise ratio limit to 150 is shown on each plot as red curves. 
As can be seen, the change is minimal.

\subsection{The metallicity, [$\alpha$/Fe] and [$\alpha$/H]  distribution functions}\label{sec:innerdiskMDF}

\begin{figure*}
\includegraphics[width=18cm]{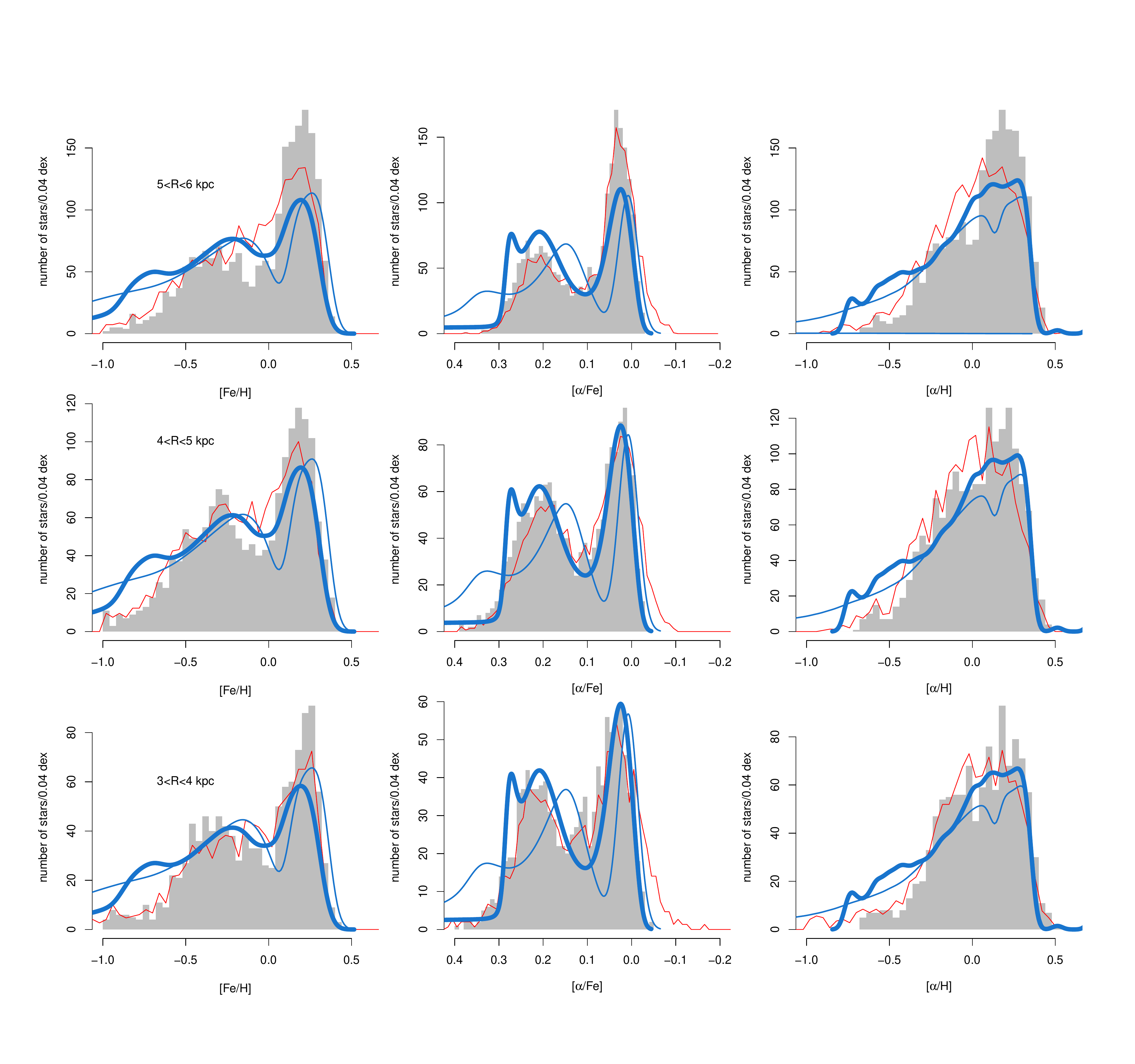}
\caption{Comparisons between the model and APOGEE data in different radial bins, as indicated on the left plots. 
From left to right: MDF, [$\alpha$/Fe]-DF, and [$\alpha$/H]-DF. 
The APOGEE data with and without the selection in the  [Fe/H]-[$\alpha$/Fe] plane are shown as
gray histograms and red curves. Red curves are normalized to the gray histograms.
The blue thick curve is our fiducial model, the blue thin curve is the smoothed model. 
Both models are normalized to have the same number of stars as the data.
} 
\label{fig:APOGEEmdf}
\end{figure*}

Fig. \ref{fig:APOGEEmdf} shows the comparison between the APOGEE distribution of metallicity (left column),  [$\alpha$/Fe] (middle) and  [$\alpha$/H] (right) of the selected objects together with the predictions of the model shown on Fig. \ref{fig:modeldist} in three different radial bins.

The first thing to note is the difference in the shape of the observed MDF in the inner disk and at the solar vicinity 
(and the difference is visible starting already at R$\lesssim$7kpc). The former reflects the bimodality seen in the 
[$\alpha$/Fe] distribution, which we discussed in \cite{haywood2016a} and is also clearly seen on the [Fe/H]-[$\alpha$/Fe] distributions 
(lower plots of Fig. \ref{fig:mdfinnerdisk}),
while the latter peaks at solar metallicity and is unimodal \citep[e.g.][]{haywood2001}.
The MDFs in the inner disk are clearly bimodal, with peaks at metallicities [Fe/H]$\sim$-0.30
and +0.25 dex, corresponding respectively at the thick disk and the thin (inner) disk, two populations that are 
in minority at the solar radius, while there is clearly a dip at solar metallicity. 
The rise of these two populations in the inner disk is well in line with 
the small scale lengths measured by \cite{bovy2016} on the APOGEE data for the alpha-enhanced and 
the metal-rich populations.

Our model is also shown on each plot, being normalized to have the same number of stars as in the observations.
We emphasize that these are not fits, but are entirely derived from the SFH deduced 
by fitting the model to the age-[Si/Fe] of the inner disk as observed in the solar vicinity, so there is no
adjustment whatsoever on any radially dependent parameter (radial migration, infall timescale, etc).
For this reason, we consider that our fiducial CBM is a satisfactory first representation of the data,
capturing the main features of all three distributions in all three distance bins: the width of the distribtions, the position 
of the two main peaks, and the presence of the dip in the MDFs and alpha-DFs.
Obviously, there is also room for ajustment. For example, the errors in the abundances are uncertain. While 
we convolved our models with an uncertainty of 0.05 dex on metallicities, using the estimate from \cite{bovy2016}, 
individual internal errors in the catalogue are peaked at about 0.02 dex. 
Second, the bump  at [Fe/H]$\sim$-0.8 dex in our model is due to a small variation in the
SFH and is not significant, but it remains however that 
both our fiducial and the smoothed models are overpredicting the number of stars below this metallicity.
There may be several reasons that can explain this offset.
Firstly, giant tracers favor younger stars and possibly underestimate 
older (alpha-rich) objects, see \cite{bovy2014}. We do not expect this bias to affect
only stars at [Fe/H]$<$-0.7 dex, because all stars older than 4-5 Gyr may be affected, and 
most significantly thick disk objects. However, if we had normalized our fiducial model on the metal-rich peak
(instead of the total number of stars), it would be apparent then that the model slightly overpredicts 
the number of thick disk stars ([Fe/H]$<$0 dex), and this could be explained by the bias induced by giant tracers. 

The second reason that comes to mind is the fact that the 'instantaneous accretion' model -- that is the closed-box model --
is a (too) rough approximation of a situation where accretion occurs rapidly. 
Star at [Fe/H]<-0.7~dex have ages > 12 Gyr. It is very well possible that a residual ‘G-dwarf problem’ remains at these metallicities/ages.
If the accretion was fast (but not instantaneous, as modeled here), the number of stars at ages > 12 Gyr and [Fe/H]<-0.7 dex 
would be less than predicted as in the standard 'G-dwarf problem', shifted to lower metallicities. 
Taking into account a small (but not zero) accretion time at ages $>$ 12 Gyr would probably improve the 
comparison between the models and the data, in a similar way standard infall models are ‘solving’ the G-dwarf 
problem at higher metallicities.

The comparisons between the APOGEE [$\alpha$/Fe] distribution function (DF) and our standard model were already shown in \cite{haywood2016a}, 
we add here our smoothed model. 
The comparisons between the APOGEE [$\alpha$/H]-DF and the models are given on the right column. Because iron is absent from
these plots, the comparisons are independent of the modeling of the iron production and in particular the time-delay
function used for the SNIa. 

To conclude, main features of the MDF,  [$\alpha$/Fe]-DF and [$\alpha$/H]-DF are  reproduced by our model with the exception of the offset at the metal-poor end as previously discussed.

We emphasize that the aim here is not to fine-tune the comparisons and to design a best fit model, but to check how, 
in first approximation, a CBM fitted on inner disk stars found in the solar vicinity, is capable of representing
the main observed features of chemical abundances.
Reproducing the details of observed distributions is also less meaningful given that the APOGEE data has its own 
uncertainties (contamination by outer disk solar vicinity stars and the distances, among others), 
while the MDF as a function of R is likely to improve significantly in the near future, 
with Gaia DR2 in particular.

\subsection{The age-metallicity relation}

\begin{figure}
\includegraphics[trim=50 80 50 80,clip,width=9.cm]{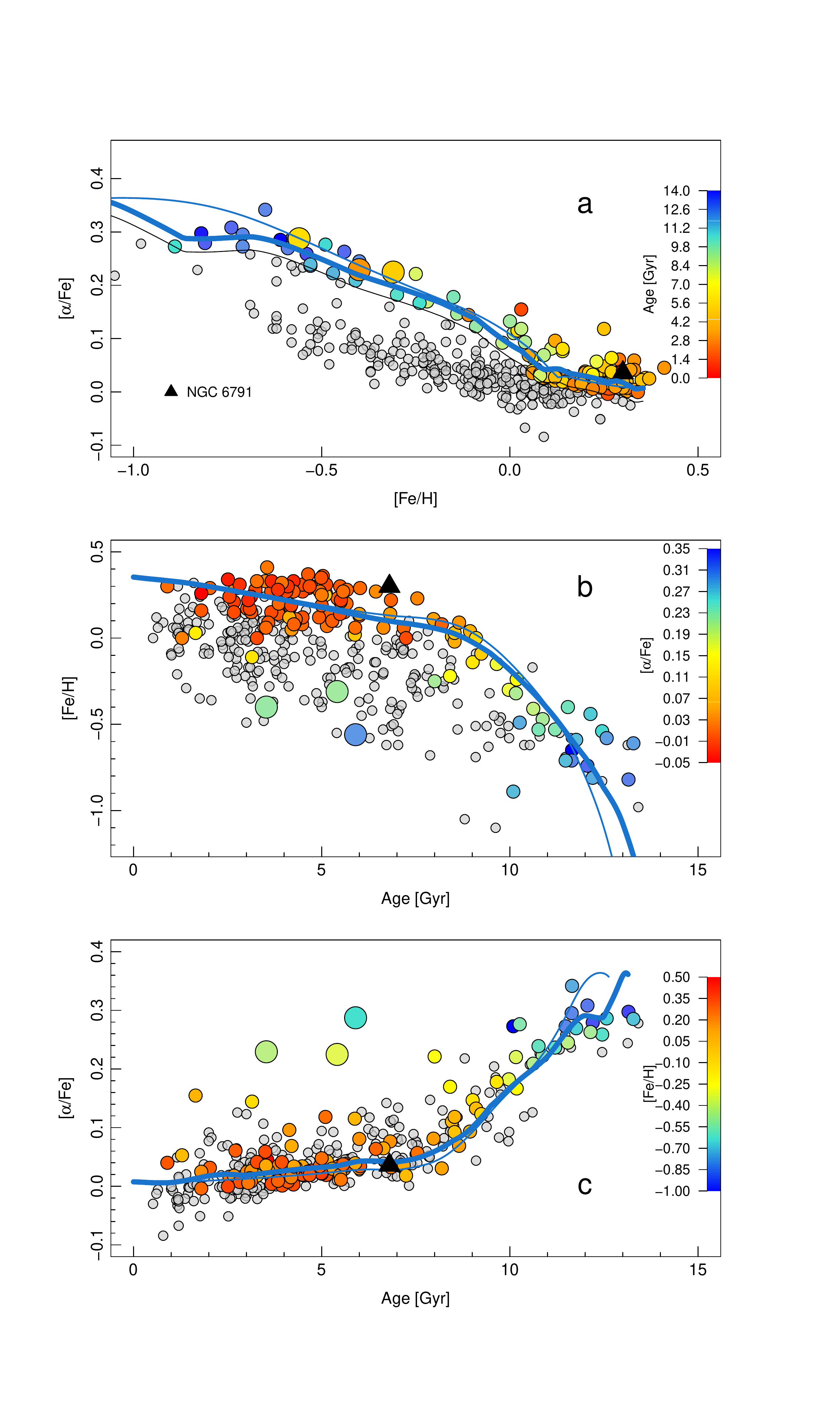}
\caption{
(a) Selected inner disk stars from the sample of \cite{adibekyan2012}, together with our models (blue curves).
The stars are selected above the fiducial model lowered down in [$\alpha$/Fe] by 0.025 dex (black curve on plot (a))
in the [Fe/H]-[$\alpha$/Fe] distribution. The color scale codes the age of the stars from \cite{haywood2013}.
(b) The age-metallicity distribution of inner disk stars, described in the text.
The observed distribution widens at younger age ($<$5 Gyr) because of the increasing difficulty to separate the 
inner disk sequence from the OLR (local) stars at lower alphas. 
(c) The age-[$\alpha$/Fe] relation for the selected stars, together with the model.
The three points with larger symbols are apparently young, alpha rich objects. 
The black triangles indicate the position of NGC 6791 at age=6.8 Gyr, [Fe/H]=+0.3 and [$\alpha$/Fe]=+0.036
(see text for references).
}
\label{fig:innerdisk}
\end{figure}

The most difficult constraints to obtain at the moment are those depending on ages, 
and are still limited to the strict solar vicinity. 
Our only possibility is therefore to select ``inner disk'' stars from a solar vicinity sample, 
and we use the sample of Adibekyan for which we have age measurements (Haywood et al. 2013, see also Haywood et al. 2015). 
Fig. \ref{fig:innerdisk}(a)  shows the sample of stars with ages from \cite{haywood2013} and 
metallicities from \cite{adibekyan2012}.
To select inner disk stars more accurately, we keep the 
stars that are above the black curve in plot (a), which is the model representing
the data (blue curve) lowered by 0.025 dex. We impose a stricter limit than for stars at R $<$ 6 kpc, because
the contamitation at the solar vicinity is likely to be more important.
This selection is likely to be accurate for the part of the inner disk sequence well
separated from the local stars, which is true for [$\alpha$/Fe]$>$+0.05 dex, or ages$>$7 Gyr. 
Below this limit the contamination by local stars will increase.
For the stars selected as inner disk objects, color codes the age according to the vertical scale on the right of the plot.
The three larger circles represent the alpha-rich, young stars identified in our sample \citep{haywood2013, haywood2015}, while 
the black triangle represents the position of the open cluster NGC 6791 in all three plots. 

Plot (b) shows the age-metallicity relation for the whole sample and for the inner disk stars.
The dispersion increases a bit at age $<$5 Gyr because our selection of ``inner disk'' stars
at [$\alpha$/Fe]$<$0.05 dex most likely includes ``OLR'' stars, which have lower metallicities at a given age.
At even lower metallicities ([Fe/H]$<$-0.2 dex) and age less than 7 Gyr, 
the three young alpha-rich stars are also visible.
Otherwise, we note that the age-metallicity relation is 
in good agreement with the model. 

Fig. \ref{fig:innerdisk} shows two important results: (1) when inner disk stars are selected, the dispersions in chemical 
abundances and metallicities at a given age are very small, both on the age-metallicity and age-[$\alpha$/Fe] plots. It means that, contrary to what is observed 
at the solar vicinity, where the spread in metallicity at all ages is observed, {\it samples of in-situ inner disk 
stars should show tight age-chemistry relations, or small dispersion at all ages. }
This point is discussed further in section \ref{sec:thesame}.
(2) The second result is that the evolution of the inner thin disk (ages $<$ 7 Gyr) is in continuity with that of the thick disk,
and corresponds to the upper envelope of the age-metallicity distribution observed at the solar vicinity. 
Within the OLR, the thin disk started to form stars from the conditions left by the thick disk  after the quenching 
episode, and continued a monotonic enrichment, illustrated by the model, with no apparent dilution.

The OLR, by establishing a barrier limiting the inflow of gas in the inner disk, presumably permitted 
a monotonic enrichment described by the CBM. At the OLR and beyond, the ISM left by 
the thick disk evolution may have mixed with lower metallicity gas present in the outer disk, 
explaining why stars at any given age at the OLR and beyond have a metallicity lower than inner disk stars.
As mentioned in Haywood et al. (2013), the lower metallicity gas coming from the outer disk could have been 
polluted by metals during the thick disk phase.
\\

To summarize, we find that our model is a good representation of the age-[$\alpha$/Fe], [Fe/H]-[$\alpha$/Fe], age-metallicity, and metallicity 
distributions of the inner disk. \cite{haywood2016a} also have shown that the model reproduces satisfactorily the [$\alpha$/Fe] distributions of the APOGEE 
survey in the distance range considered here. To our knowledge, it is the first time that a Galactic chemical evolution model provides a good match to such a large range of observational constraints of the inner disk.\\

{\it The case of NGC 6791}\\

Also added to Fig. \ref{fig:innerdisk} are the positions of the metal-rich cluster NGC 6791. 
This cluster is often presented as an outlier to the chemical evolution of the disk 
\citep{carraro2006, geisler2012, carraro2014}.
However, with age estimates between 6.8$\pm$0.4 and 8.6$\pm$0.5 Gyr (Basu et al. 2011), a metallicity of $\sim$+0.3 dex (+0.3$\pm$0.02 dex, Boesgaard et al. 2014, +0.29 dex, Brosgaard et al. 2011), NGC 6791 is typical of the evolution 
of the inner disk, as described here.
The alpha element abundance of NGC 6791, as deduced from the study of \cite{boesgaard2015}
is +0.036 (taking the mean of Mg, Si and Ti), and falls well on both the observed age-[$\alpha$/Fe] and [Fe/H]-[$\alpha$/Fe] relations, 
see Fig. \ref{fig:innerdisk}.
If the age of NGC6791 is confirmed to be greater than 7 Gyr, it implies that it would have formed 
during the quenching episode that occured in the MW at this time  \citep{haywood2016a}. Our SFH, taken 
as a birth probability function (and assuming that our age scale is correct), and the age-metallicity relation,
favor a slightly lower age determination.

\cite{jilkova2012} tried to reconcile the present orbit of NGC 6791 
(R$_{peri}$=5-5.3kpc, R$_{apo}$=8.5-9.1kpc, and eccentricity in the range 0.23-0.29, with the eccentricity defined 
as (R$_{apo}$ - R$_{peri}$)/(R$_{apo}$ + R$_{peri}$) with a scenario where the cluster originated 
in the inner disk (R$<$5kpc), possibly down to R$\sim$3 kpc, and then moved outwards by radial migration. With the properties of the inner disk 
described here however, the cluster does not have 
to originate from Galactocentric radii as small as 3~kpc. If the cluster was born within 6-7 kpc
from the Galactic center, it would still be  fitted to its local environment since 
the mean metallicity of the thin disk in this radial bin is about +0.25 dex.
NGC 6791 is the best example of a metal-rich object which can pollute the solar orbit by having eccentricities
sufficiently high that blurring only is enough to explain their presence at the solar circle.
The study of \cite{jilkova2012} is also a nice illustration of the result found by \cite{halle2015}, 
that the OLR of the bar plays the role of a barrier for the migration of stars. 
The Fig. 4 of \cite{jilkova2012} shows that this is true not only when the bar is the only asymetry, but also 
when there are spiral arms, or both a bar and spiral arms. In all cases, the cluster has a negligible probability
to cross the OLR of the bar. 
Note that this effect is also nicely confirmed by \cite{monari2016}, who also show that the OLR acts as a strict barrier 
to the stars when the bar is the sole asymetric perturbation (their Fig. 11, left plot), while their 
simulation shows that even when there is a coupling between the bar and the spiral arms, the extent
of radial mixing is extremely limited (less than 500~pc for most of the stars, less than 1kpc for essentially
all the stars).

\begin{figure}
\includegraphics[trim=80 100 40 140,clip,width=10.cm]{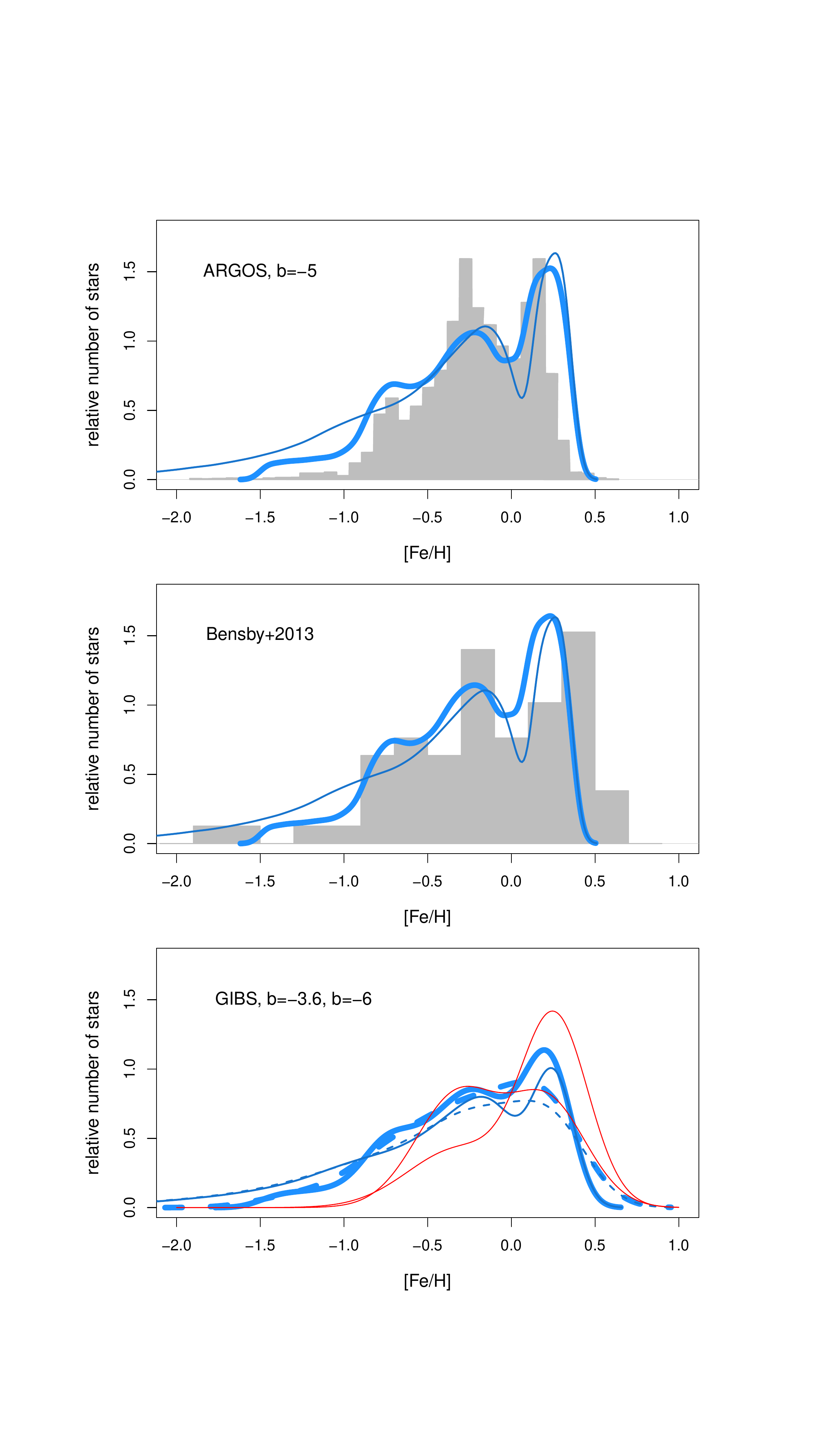}
\caption{Comparison of the fiducial model MDF (thick blue line) and smoothed model (thin blue line) with different observed bulge MDFs from Ness et al. (2013, top), Bensby et al. (2013, middle) and \cite{zoccali2017}. Two MDFs at different latitudes \cite{zoccali2017} are shown, b=-3.6 and -6$^\circ$, and in this case models have been convolved with a gaussian of 0.1 dex (continuous blue curves) and 0.2 dex (dashed blue curves).}
\label{fig:mdfbulge}
\end{figure}

\begin{figure}
\includegraphics[width=9.cm]{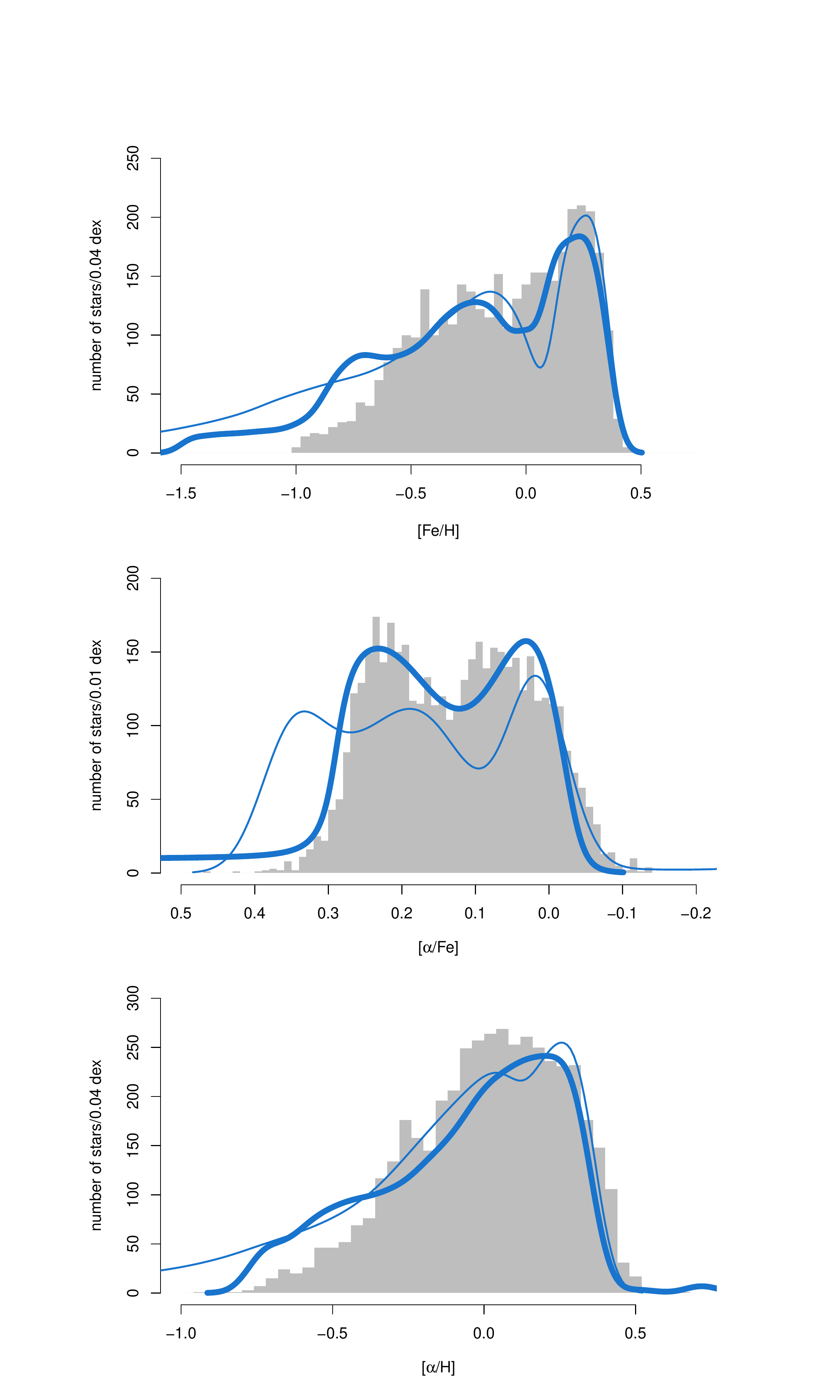}
\caption{APOGEE sample of stars within 3 kpc from the Galactic center, chosen to represent the bulge population, 
compared with our models. The models are normalised to have the same number of stars as the data in the metallicity range [Fe/H]=-1 to +0.5 dex.
Thin curve: smoothed model. Thick curve: fiducial model.
Upper plot: comparison of the metallicity distributions. Middle plot: comparison of the [$\alpha$/Fe] distributions. Lower plot: [$\alpha$/H] distributions.
}
\label{fig:bulgeapogeerc}
\end{figure}

\section{Results: Chemical  evolution of the bulge}\label{sec:bulge}

If the majority of the bulge is made of material transferred from the disk,
we  expect the stars of the bulge to have stellar ages and chemical characteristics (both the global MDF and abudance ratios) 
not different from that of the disks. 
The chemical evolution of the MW bulge is commonly modeled as an independent system, even when it is acknowledged that part of its structure
and content must have come from the disk through dynamical instabilities \citep[e.g.][]{grieco2012,tsujimoto2012}.
This is because of two  properties, often thought to be the landmark of the bulge:

(1) the MDF of the bulge is very broad, reaching both high and low metallicities, and has a minimum at solar 
metallicity, while the disk MDF, as sampled in the solar vicinity, is narrower and peaks at solar metallicity. 
This is often presented as the evidence that the bulge origin is not in the disk (see e.g, McWilliam, 2016, Fig. 3). 
We argue below that since only stars whose guiding radii are within the OLR participate to the bar, comparing the 
bulge and solar vicinity MDFs is incorrect, while, as we show below, the MDF of the inner disk and the MDF of the bulge are compatible.

(2) The bulge seems to be uniformly old ($>$10 Gyr). Reaching the high metallicities ([Fe/H]>+0.4 dex) observed in the bulge within 3-4 Gyr after the Big-Bang is challenging, and  requires a particular chemical evolution.
Recent works however suggest that the bulge is not uniformly old. \cite{bensby2011,bensby2013} 
show that the age spread in the metal rich component of the disk is at least $\sim$10 Gyr, 
with stars that have ages as low as 2 Gyr (see also \cite{schultheis2017, haywood2016b,catchpole2016, groenewegen2005}).

Allowing the bulge to have much younger stars than previously thought releases the difficulty of having 
to design models capable of reaching high metallicities within a few Gyr. 

We now explore in more details how our model matches the bulge chemical characteristics.

\subsection{The bulge MDF} \label{sec:bulgeMDF}

Perhaps the most significant hurdle in identifying the bulge population with the thick and thin disks is the MDF of the bulge, which is much broader than the disk MDF measured at the solar vicinity. 
As said in the introduction, stars are driven to the inner regions to form the bar through dynamical instabilities from regions inside the OLR, 
as quantified in detail in \cite{dimatteo2014}, \cite{halle2015} and \cite{halle2018}. In the MW, the OLR resonance is located 
near the solar circle (7-10 kpc) according to \cite{dehnen2000} and \cite{perez2017}.
It implies that, if the bulge is essentially made of disk stars, it must be dominated by inner disk (R$\lesssim$ 7kpc) objects, 
and not by the kind of thin disk stars that dominate the solar vicinity and beyond. \cite{dimatteo2014,dimatteo2015} explicitly  
identified the population A, B and C of Ness et al. (2013) as the thick disk (for C and B) 
and the inner thin disk (for population A). 
\cite{fragkoudi2017} showed that this scenario is able to reproduce the mean metallicity map of the bulge, as revealed by APOGEE.
We thus expect the (global) MDF of the bulge to be the same as the MDF of the inner disk.
There are two caveats here. The first is that we lack a measurement of the global MDF of the bulge,  having only estimates of the MDF in different 
limited regions. Hence, one must make a choice to select a distribution representative of the global MDF.
Second, because stars are redistributed in the bulge according to their initial kinematics (i.e thick or thin disks) 
and initial location in the disk, the MDF will vary as a function of latitude and longitude. 

For instance, Di Matteo et al. (2014) have found that stars originating closer to the OLR tend to populate the outer region of 
the bulge. 
Most importantly, the relative fraction of thin and thick disk stars is a strong function of latitude, with the 
metal-rich thin disk dominating at low latitude and the metal-poorer thick disk at high latitude (except along the bulge minor axis, at very low latitudes, where this trend seems to invert, see Zoccali et al 2017).
Hence, we do not expect the observed MDF sampled at any given location in the 
bulge to be closely similar to the disk MDF. Since we do not model the spatial distribution of these two populations in the 
bulge, the comparison of the model and observed MDF can only be qualitative, 
with attention to the most important features rather than to details. We refer to Fragkoudi et al. (2018, submitted) for a detailed
modelling of the MDF of the bulge as a function of longitude and latitude.

Fig. \ref{fig:mdfbulge} (top) shows the MDF of our model (blue curve) together with the MDF from the ARGOS survey at  b=-5$^{\circ}$ \citep{ness2013b} and with the MDF from \cite{bensby2013} (middle plot). The distribution from \cite{bensby2013} is made of stars that are mostly between latitudes $-6^{\circ}<b<-2^{\circ}$ and longitudes $-6^{\circ}<l<7^{\circ}$, while the MDF from \cite{ness2013b} is made of a series of fields at different longitudes, in the interval $-15^{\circ} \le l \le +15^{\circ}$ but limited to $b=-5^{\circ}$. 
Also, the MDF from \cite{ness2013b} contains several thousand stars, while the sample from \cite{bensby2013} contains 58 dwarfs and subgiants stars observed through microlensing events. 
While observed MDFs in the galactic plane would favor the contribution of the thin disk (the metal-rich peak), those out of the plane favor the thick disk (the metal-poorer peak). Hence, the field from \cite{ness2013b} at b=-5$^\circ$ is unlikely to severely underestimate one of the two, but however systematic differences with the model MDF are not unlikely, and the sampling at this particular latitude may not represent the ratio of the thin and thick disks of the model -- that is the observed ratio between the metal-rich to metal-poor peak. 

In spite of these differences, the two agree fairly well, and in particular they both show the dip in the MDF at [Fe/H]=0 dex.
The difference between Ness et al. and Bensby et al. at supersolar metallicities illustrates that the metallicity scale, in particular at [Fe/H]$>$0 dex, is not secure, and can vary significantly from one study to the other\footnote{As another example, Rojas-Arriagada et al. (2014) have found a systematic offset between their metallicities and those of \cite{hill2011} which amounts to +0.21 dex, for a subset of a hundred stars common to both studies.
Once the MDF from Hill et al. (2011)  is shifted to lower metallicities by this amount, the agreement between
our model and their MDF is also satisfactory, see \cite{haywood2016b}.}.
Bottom plot compares the models with MDFs from \cite{zoccali2017} at two latitudes, -3.5$^\circ$ and -6$^\circ$ and which combine a range of longitude (from -8 to +8$^\circ$), using the gaussian components given in their table 3. Each MDF combines samples 
at different longtitudes (their Fig.7). These two latitudes were chosen for the same reason that we chose ARGOS field at $b=-5^\circ$: the spatial segragation at theses latitudes will not favor too strongly either metal-poor or metal-rich population. 
According to \cite{zoccali2017}, the errors in metallicities range from 0.1 to 0.4 dex (from metal-poor to metal-rich stars), with a mean at 0.2 dex.
To take this into account, we convolved our standard model with gaussian with two dispersions: 0.1 and 0.2 dex, represented by the thick continuous 
and dashed curves. The two observed MDFs illustrate the significant variability between two MDFs separated by only $\sim$ 350 pc in the vertical direction, with the metal-rich component representing 67\% of the stars in the field at 3$^\circ$, and 20\% less at 6$^\circ$ \citep{zoccali2017}.

Figure \ref{fig:bulgeapogeerc} shows the comparisons with the APOGEE survey. The stars from APOGEE representing the bulge have
been selected within 3~kpc from the Galactic center (longitudes between 0 and 20$^\circ$, latitudes between -15 and +15$^\circ$) with the same criteria as for the inner disk.
The same offset between the data and models at low metallicities observed for the inner disk is seen on the bulge APOGEE data, with the number of stars overpredicted by the models at [Fe/H]$<$-0.7 dex.

The dip is not visible in the MDF of the APOGEE data (Fig. \ref{fig:bulgeapogeerc}, top plot), but the overall 
distribution is nonetheless well reproduced by the model at [Fe/H]$>$-0.5 dex. A reason for the absence of the dip may be due to the distance distribution of APOGEE stars in the Galactic bulge, which is biased towards the near side of the bulge (see, for example, Fig 2 in Ness et al 2016), and against the lower metallicity stars. A discussion of this distance bias and its impact on the bulge MDF is discussed in Fragkoudi et al, submitted. 

In several other datasets, the bimodality of the bulge MDF is apparent.
\cite{gonzalez2015} found two components centred on [Fe/H]=-0.31 and +0.26 dex.
\cite{uttenthaler2012} found a transition at [Fe/H]$\sim$0 dex (their Fig. 8), and two peaks at [Fe/H]=-0.57 and +0.30 dex, although their field, being at b=-10$^{\circ}$, shows a much less prominent metal-rich peak. Zoccali et al. (2017) analyse 26 different fields and find a transition between two main peaks at about solar metallicity in most fields. 
As for the inner disk, in our model, the dip in the MDF corresponds to the lull in the star formation at $\sim$8 Gyr marking the transition from the thick to the thin disks. 

Overall, we consider the match between the model and the data to be satisfactory given that (1) the model is entirely designed by fitting chemical abundances of stars of the inner disk sequence observed at the solar vicinity, and no tuning has been introduced to fit the inner regions. 
(2) the uncertainties that still exist in the metallicity scales, and the apparent variability between fields at the same latitude \citep{zoccali2017}
(3) the difference in sampling between the four different datasets considered here.

\subsection{The bulge age-metallicity relation}

\begin{figure}
\includegraphics[trim=80 40 40 20,clip,width=10.cm]{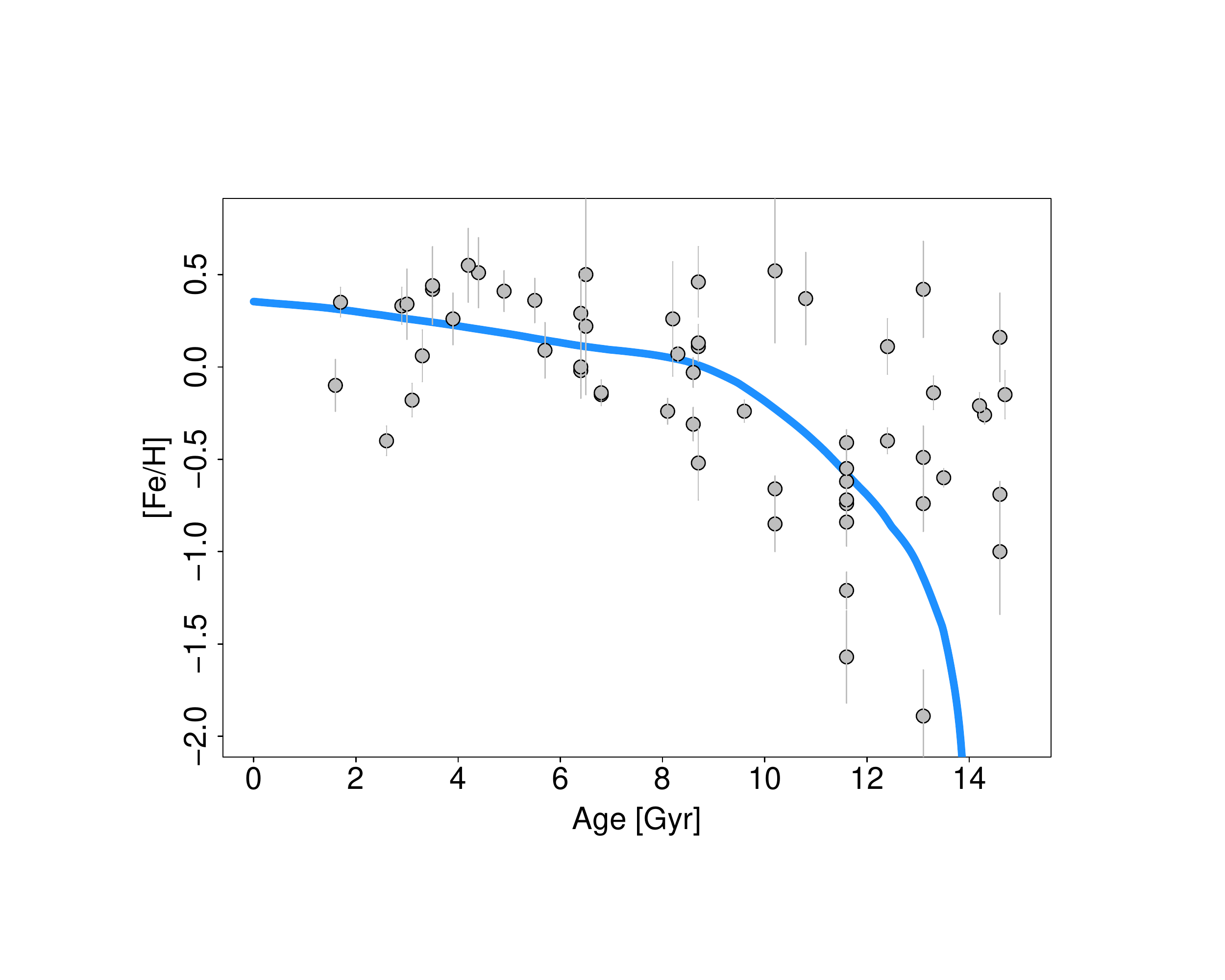}
\caption{
The age-metallicity distribution of the bulge from Bensby et al. (2013) compared to our models.
}
\label{fig:bulgeamr}
\end{figure}

Recent evidences that the bulge also contains young stars come from \cite{bensby2013}, who have shown that stars with age $<$ 8 Gyr exist in the bulge, confirming other previous findings (see for example \cite{vanloon2003}, and references therein).
Fig. \ref{fig:bulgeamr} shows the comparison between our model AMR and the bulge data from \cite{bensby2013}.
Obviously, the ages have large uncertainties, but the model is nonetheless a good representation of the data, with no 
obvious disagreement.

\subsection{Alpha abundances}

Because the high-alpha abundances constitute also an age sequence \citep{haywood2013}, it is interesting to compare our model to the [Fe/H]-[$\alpha$/Fe] data and to the alpha element distribution. 

Figure \ref{fig:bulgeapogeerc} (middle \& lower plots) shows the alpha element distribution function (both as [$\alpha/Fe$] and [$\alpha$/H]) 
of APOGEE stars compared to our models.
As in \cite{haywood2016a} on the inner disk data, the fiducial model is a good representation of the bulge  [$\alpha/Fe$] distribution. 
The smoothed model shows less good agreement, the reason being the more prominent tail of stars at metallicities lower than -0.7 dex, 
and which produces a third bump at [$\alpha$/Fe]$\sim$+0.35.
The comments made about the APOGEE MDF also apply to the [$\alpha$/H] distribution: the comparisons are overall 
satisfactory, with a significant overproduction of stars at low [$\alpha$/H] (old objects).

\begin{figure}
\includegraphics[trim=80 40 40 20,clip,width=10.cm]{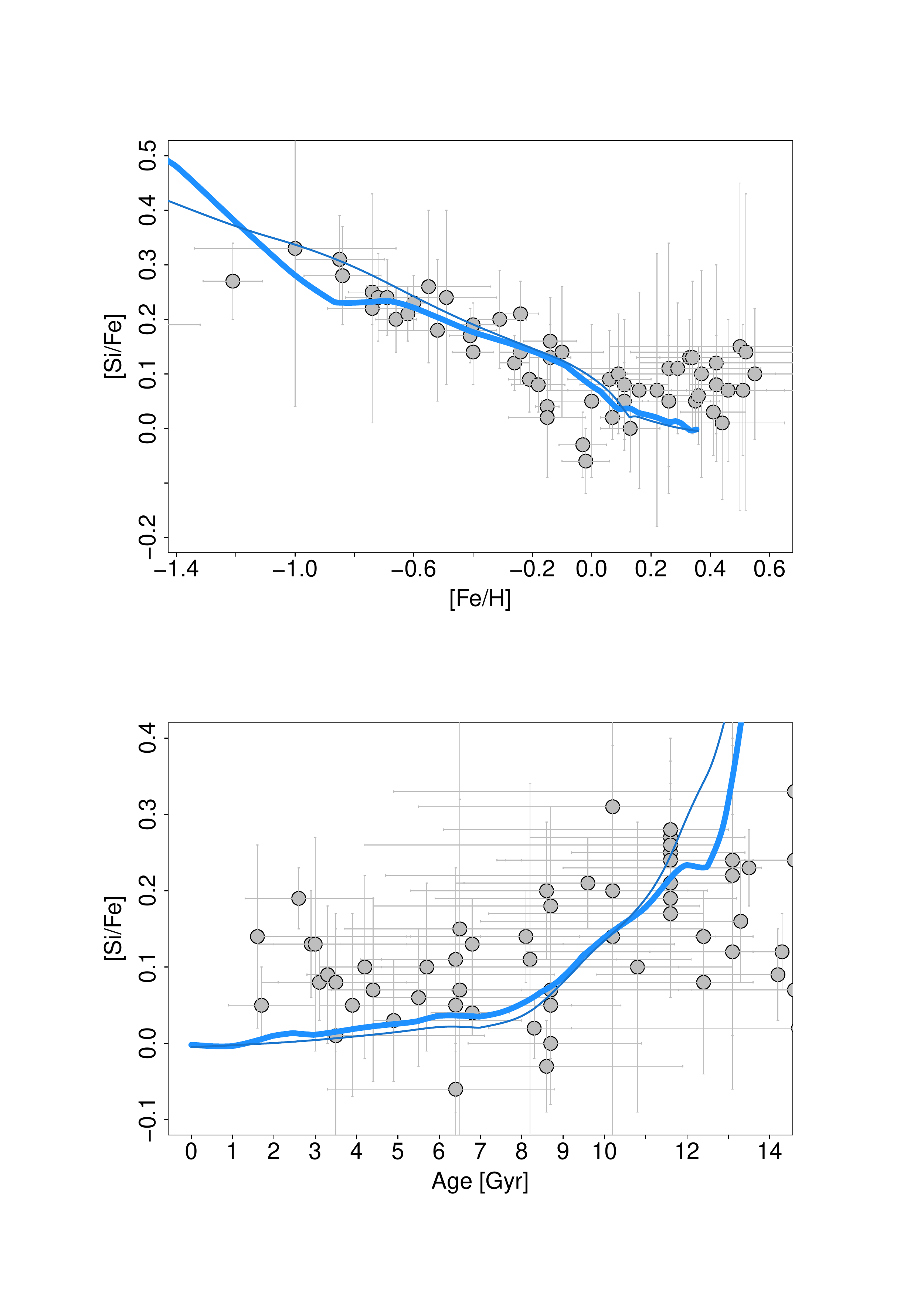}
\caption{The [Fe/H]-[Si/Fe]  and the age-[Si/Fe] distributions of the bulge 
 from Bensby et al. (2013) compared to our models (top and bottom respectively). }
\label{fig:bulgeafeh}
\end{figure}

Fig. \ref{fig:bulgeafeh} shows the silicon abundance as a function of [Fe/H] and age for the microlensed stars in \cite{bensby2013}. 
The model does not reproduce the upturn seen in \cite{bensby2013} data at super-solar metallicities, but we note that the error bars are large at these metallicities, our model still being compatible with the data within one sigma, and that this upturn is not always observed. Gonzalez et al. (2011) for instance shows no evidence of such feature in none of the alpha element they have studied, neither it is in Gonzalez et al. (2015) nor in Ryde \& Schultheis (2015).
The bottom plot shows the age-[Si/Fe] distribution of Bensby et al. (2013). As for the age-[Fe/H] distribution, the models
are compatible with the data, but the data offer no real constraint because of the large uncertainties on ages.

\subsection{Conclusion}

To conclude, our fiducial model provides a good representation of a large range 
of observational constraints of the bulge: MDF, alpha-DF, [Fe/H]-[Si/Fe] distribution, age-[Si/Fe] and age-metallicity 
distribution with the same model as the inner disk, which we find remarkable given that the model (for both the inner disk and bulge)
was built only from the comparison with inner disk stars sampled at the solar vicinity.
The model so far has no free parameters (apart from the initial assumption of a CBM), 
while in standard chemical evolution models of the bulge \cite[e.g.][]{tsujimoto2012,grieco2012}, the datasets are 
fitted allowing for several additional (and unconstrained) parameters such as the gas accretion timescale, the IMF, 
which usually differ from those of the disk, while in our case the model is the same.

\section{Discussion: The inner disk-bulge population}\label{sec:thesame}

The application of the CBM to the bulge and inner disk is valid if these can be considered as a unique system in all their characteristics, and if their evolution was not radially dependent. 
We discuss these two points in the following.

\subsection{Are the inner disk and bulge the same populations?}

If the inner disk and bulge are the same population, their kinematic, chemical, and age characteristics must be compatible.
The compatibility of the disk and bulge kinematic properties will be discussed at length in Di Matteo et al. (2018, in prep.) and will not be discussed here. We focus here on the chemical properties and on the age problem of the bulge.

\begin{itemize}

\item \emph{The MDF}\\

The possible identity of the bulge and the disk has been questionned, observationally, in particular on the basis of the strongly different MDFs of the bulge and solar vicinity, \citep[see, for example, ][]{mcwilliam2016}.

In the previous section, we explained why these two should not be compared, while we have shown that 
that there is good evidence for a similar MDF of the inner disk and bulge. This similarity appears both in the width and in the shape of the MDF, and both can be reproduced by the same chemical evolution model. It is also shown in \cite{fragkoudi2017} that the mean metallicity maps of the bulge as a function of longitude and latitude in the model are in good agreement with the APOGEE data.\\

\item \emph{The abundance ratios}\\

Because the bulge is mainly a bar made of thick and thin disks stars formed inside the OLR, as argued in Di Matteo et al. (2014), the chemical abundance patterns of the bulge and the disk inside the OLR should be similar.
Classically, chemical abundances of the solar vicinity and bulge are compared directly, while in fact the solar neighbourhood, being a transition zone between the outer and inner disks, (Haywood et al. 2013, Haywood 2014) is polluted by outer disk stars and stars formed at the OLR, which are essentially absent inside the OLR. The resulting differences, while expected, have been used to argue that the bulge and the disk have a different origin. 
This is one of the important point to take into account when comparing the disk and bulge abundances. The other crucial point are the systematics. 

To assess if the differences between the thick disk and bulge stars are real, we must understand what are the level of the systematics that are expected between different samples. 
For instance, if the systematics measured on two samples of bulge stars are of the order of the differences measured between the bulge and the local thick disk, then we may infer that these differences are not significant. Both aspects are explored in detail in Appendix A.  
We show that when these caveats are taken into account, either the match for several elements abundances is remarkable (alpha elements, Ni, Zn, Cr, Ba, Eu, La), or the differences, when they exist, are likely to remain within the level of systematics (Mg). Differences are not explained for some elements (Cu), but the usually small number of existing datasets prevents any strong conclusions. \\

\item \emph{The age of the bulge}\\

The supposedly exclusively old age of the bulge \citep{clarkson2008, valenti2013, nataf2015} is  cited has one of the main differences with the disk, the hurdle being the CMD of the bulge in the SWEEPS field observe by the HST \citep{clarkson2008}. 
\cite{haywood2016b} explored in detail how stars following the age-metallicity relation of our model 
would superpose in the CMD to give the tight turn-off that has been observed in \cite{clarkson2008}.
But the main constraint comes from the wide spread in metallicity observed in the bulge:
combined with an exclusively old bulge, this would automatically generate a CMD with a turn-off too wide compared to the observation. There is presently no other way to have a CMD compatible with the observation than to assume that a significant fraction of the stars are younger than 8 Gyr. Which means that the narrow turn-off of the CMD, which was cited as evidence of a mono-age, old bulge, is in fact the demonstration of the contrary. The metal-rich stars have to be younger than 8 Gyr to allow for the observed wide spread of metallicities to be compatible with a narrow turn-off. 
We note, as already mentioned in \cite{haywood2016b}, that the tight turn-off of the bulge seen in the SWEEPS field is also evidence that the age-metallicity relation in the inner disk and bulge must be tight, otherwise a significant spread in metallicity at a given age would produce a significant spread in color at the turn-off that is not observed. 
The most recent results from Bensby et al. (2017) confirm their previous findings (Bensby et al., 2013) and show that the fraction of stars younger than 8 Gyr is about 50\%. This is higher than the percentage we estimated in the SWEEPS field \citep[35\%,][]{haywood2016b}, but a significant fraction of the stars in the Bensby et al. (2017) dataset is observed closer to the plane, favoring the young population \cite{ness2014}.
\end{itemize}

\subsection{Three hints that the formation of the thick disk was not inside-out}\label{sec:noinsideout}

That the whole inner disk and bulge can be described by a simple CBM as shown by the good match obtained on the MDF, alpha-metallicity distributions, age-metallicity relation and age-[$\alpha$/Fe] relation, implies that no significant differentiation of the chemical evolution as a function of radius occurs within the limit of the system taken in consideration. This is at variance with the assumptions of the inside-out formation of the disk, which is commonly implemented in GCE models by parametrizing the infall timescale of gas accretion as a function of the distance to the Galactic center \citep[see e.g.][]{minchev2013, kubryk2015a, loebman2016}.
The disk is formed through long timescale infall: in Fig.~4 from \citet{kubryk2015a} the infall timescale is between 4 and 8 Gyr at 3 to 7 kpc from the Galactic center; in \citet{minchev2013} it is equal to 1.82~Gyr at  3~kpc, and equal to 6~Gyr at 7~kpc). The following points contradict the predictions of this scheme and of the inside-out formation scenario.

\begin{figure}
\includegraphics[trim=40 80 40 40,clip,width=10.cm]{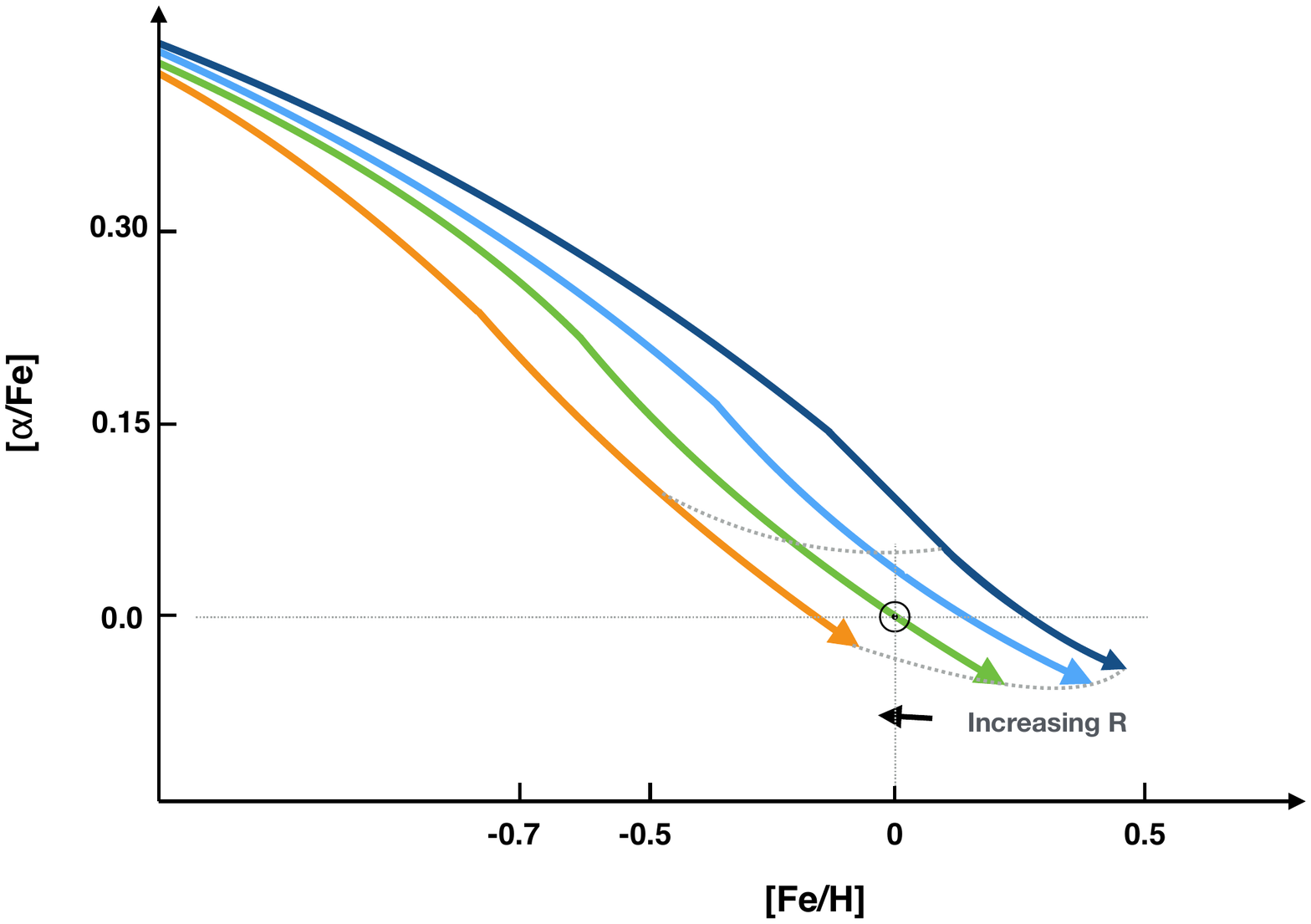}
\includegraphics[trim=40 80 40 40,clip,width=10.cm]{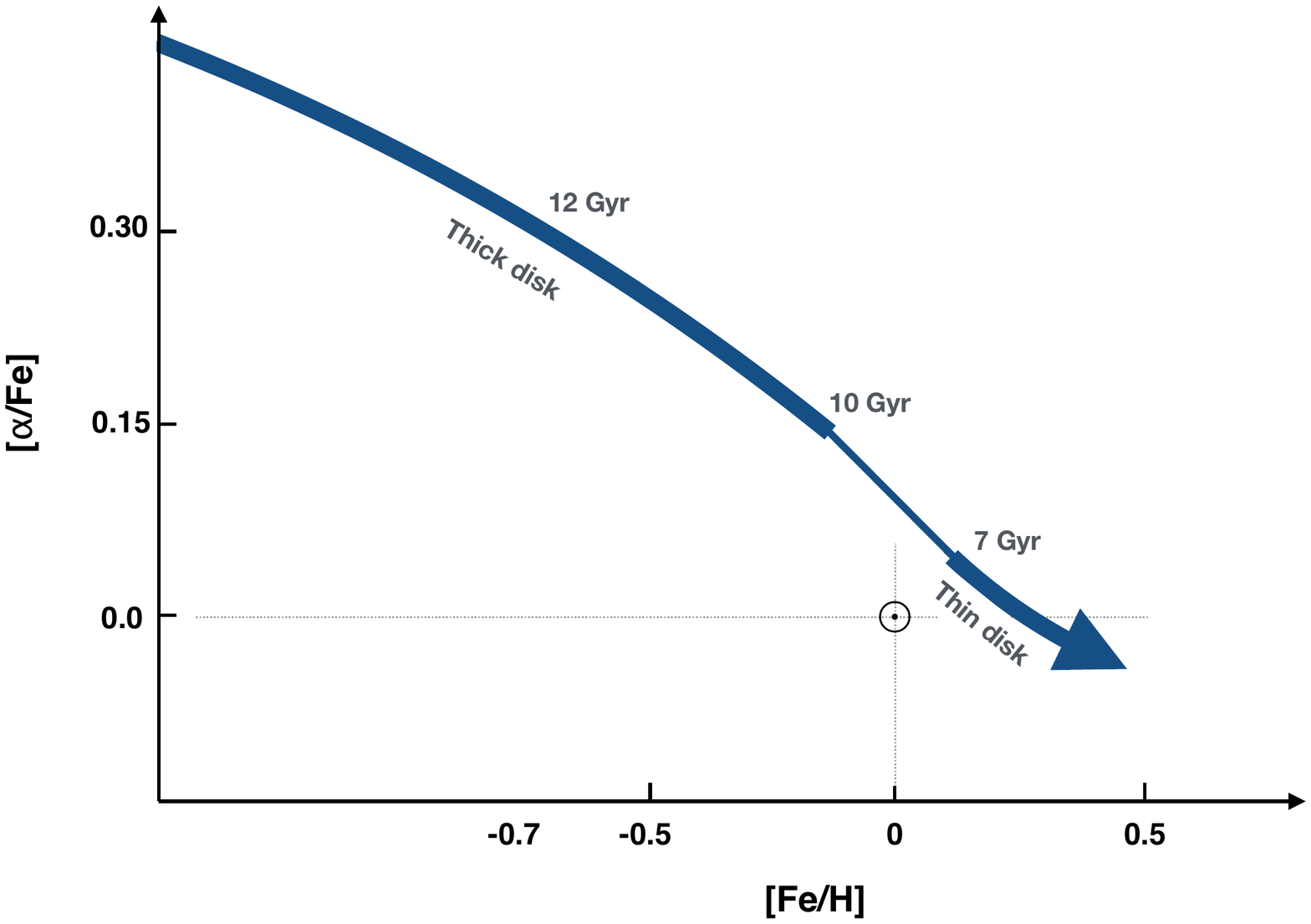}
\caption{Two different illustrative sketches of the inner disk [Fe/H]-[$\alpha$/Fe] distribution: (top) in case of the inside-out scenario,
and (bottom) for the scenario presented here. We emphasize that these schemes are meant to represent the evolution of the inner disk
only (not the whole disk). The position of the Sun (which, in the second scenario, does not belong to the inner disk evolution) is represented on each plot, together with indicative ages along the 
chemical track in the bottom plot. On this plot, the thinner 
segment (between 10 and 7 Gyr) corresponds to the quenching episode of star formation found in  \cite{haywood2016a}.
In the first scenario, a spread in metallicity and alpha abundance is expected at a given age in the inner disk.
}
\label{fig:alphafeh_sketch}
\end{figure}

\begin{itemize}

\item {\it The disk scale length does not increase with time.}
Inside-out formation scenario implies that disks grow as a function of time, which should manifests itself by the increase in time of their scale length \citep[see, e.g.,][their Fig.~1]{kubryk2015a}).
In our model, no subsantial increase of the inner disk scale length is expected  \citep{haywood2015}, because the model is a closed-box, and no inside-out process operates. 
We insist that we are discussing here the disk inside R$\lesssim$6-7~kpc, possibly corresponding to the region inside the bar OLR.
Using the SEGUE survey, Bovy et al. (2012) showed that the scale length of the inner disk stayed constant with time. This is remarquably confirmed with the APOGEE survey \citep{bovy2016}, which shows that stars that are on the inner disk sequence (high-$\alpha$ sequence in Bovy et al. 2014 paper) have remarkably constant density profiles (their Fig. 11).
The most metal-rich mono-abundance group in the work of Bovy et al. (2016) (Fig. 11 lower plot) has [Fe/H]$\sim$+0.3 dex, which is not the terminal metallicity of the solar vicinity, but the metallicity typical of the inner thin disk. 
Its scale length is 1.67~kpc when fitted with a single exponential over the whole distance range from 4 to 14~kpc, but 1.25~kpc only if fitted with a broken-exponential, which does not point to an increase in the scale length, even in the thin inner disk. 
The result of Bovy et al. (2016) clearly demonstrates that the scale length of the thick disk shows no increase. This is in agreement with the observation of MW-type galaxies at redshift $>$ 1 (when the thin disk was not yet in place) which shows also that their growth is self-similar, not inside-out (van Dokkum et al. 2013; Morishita et al. 2015).

\item {\it The radial metallicity gradient in the thick disk is flat.} 
In the inside-out formation scenario, the gradient at the end of the thick disk phase is expected to be steep. In fact, typical predicted gradients are of the order of -0.125 dex/kpc (see Fig. 4 in \cite{kubryk2015a}, Fig. 2 in \cite{minchev2013}). 
In contrast, we considered  \citep{haywood2013} that the tight age-metallicity relation in the thick disk implies that the ISM during the thick disk formation is essentially well mixed, because of the high turbulence that existed at this epoch \citep{elmegreen2006,forster2006}. These are also the conditions of our model, where the ISM is assumed to be  well mixed. 
This is suppported by the analysis of the SEGUE survey by \cite{cheng2012}, who find that essentially no radial gradient is visible on thick disk stars, as defined by stars between 1 and 1.5~kpc from the Galactic plane.
The lack of a gradient in the thick disk is not due to the effect of radial mixing.
We emphasized in  \cite{haywood2013} that mixing (by either churning or blurring) of an existing gradient would generate a dispersion in the AMR which is not observed.
If the  disk had uniform chemical characteristics, independent of radius, at the end of the thick disk formation, it is possible that the subsequent formation of the thin disk may have led also to a flat metallicity distribution in this population. We emphasize that existing measurements \citep[e.g.][]{hayden2015,bovy2016} find a steep radial gradient in the thin disk but are restricted to within $\pm$2 kpc from the Sun radius, essentially excluding the distance range considered here (R$\lesssim$ 6~kpc). In fact, Andrievski et al. (2016) suggest that the gradient may be flat within R$<$6 kpc in the thin disk.

\item {\it Chemical abundances have a very small dispersion at a given age in the thick disk.}
In the inside-out scenario, at early times, the SFR is more intense in the inner parts of the disk, producing a negative radial gradient of alpha abundances (alpha abundances decreasing with increasing radius) and metallicities (chemical evolution being faster in the inner part, the metallicity is higher in the inner disk at a given age).
Because these stars are now on significantly eccentric orbits, a spread of alpha abundances would be expected for stars of any given age at all radii.  
A significant dispersion is also predicted by these models in metallicity and [$\alpha$/Fe] as a function of age, and the same is expected in the [Fe/H]-[$\alpha$/Fe] plane. 
Fig. \ref{fig:alphafeh_sketch} illustrates the chemical tracks that may be expected in this plane:
in the case of the inside-out formation of the thick disk (plot a), the inner thick disk reaches higher metallicities at a given [$\alpha$/Fe], while due to a slower evolution, the outer thick disk would reach the same metallicities at later times. 
In contrast, if the thick disk forms from an ISM whose homogeneity is maintained by high turbulence, its chemical evolution will describe a unique track. 
The small dispersion in chemical characteristics at the end of the thick disk phase is exemplified by the age-[$\alpha$/Fe] relation. Even taking into account the observational uncertainties on age and alpha abundances, the observed spread is significantly
smaller than predicted by these models, see  \cite{haywood2015}.

\end{itemize}

These three observational evidences are strong hint that the thick disk formation was not inside-out, and that the inner disk and bulge can suitably be represented by a model where no spatial differentiation is introduced. It also implies that any detection of radial migration among stars within the OLR (the region where radial migration is expected to be the strongest) based on chemical characteristics will be difficult.


\section{The significance of the Closed-box model and the history of gas accretion in the Milky Way}\label{sec:accretion}

The model described in the previous sections, in spite of its simplicity, provides a fair representation of a surprisingly large range of observational constraints of the inner disk and bulge. The aim of the present section is to try to understand why.
Standard chemical evolution models follow two implicit principles: the SFH follows
the accretion history, and for this to be effective, the SFR is proportional to the gas density at all times. 
None of these are applicable to our CBM, and we must understand what it implies on the accretion 
history and on the star formation rate in the MW.

The latest paradigm of how gas is accreted on galaxies (see section 6.1) through cold flows, suggests that most of the gas may have been 
accreted rapidly (z$>$2-3, or in the first 2-3 Gyr) in the central parts of galaxies.
The CBM is different from this picture, because it describes an instantaneous accretion, where most accretion more likely occured at z>6. 
This may be the reason of the discrepancy observed at [Fe/H]$<$-0.5 dex (z$>$3 in the model). The 
model satisfactorily represents the data above [Fe/H]=-0.5 though, at epochs when accretion may still have been 
quite substantial (down to z$\sim2$, corresponding to  [Fe/H]$\sim$0 dex in the model).
This means either that, in the case of the Milky Way,  most accretion was terminated at significant earlier times (at z$>$3), or that the building of the disk between z=2 and z=3 was not dependent on the accretion rate. This is discussed in \ref{sec:coldacc}.


The second constraint that limits the CBM is that any MW model with high gas content, 
and in particular the CBM, is bound to produce unreastic age distribution if the SFR is directly dependent 
on the total amount of gas. Yet, to be valid, the model also has to be on the SK relation. We explore in Section 
\ref{sec:sklaw} at what conditions on the gas content this is verified. 
Also, a strict dependence with the SK law ignores possible
variations in the star-formation efficiency \citep[e.g.,][]{schreiber2016}.
The data for the MW, for instance, show evidence for a quenching episode that occured while the total amount of gas and its surface density in the
MW must have been was quite high \citep{haywood2016a}, if the MW evolution was similar to other 
galaxies of the same stellar mass.  This result implies that the star formation efficiency must have varied significantly at this
epoch. Such a variation in the SFH cannot be explained in standard GCE models. Standard GCE models enforce a strict relation between gas
density and the star-formation rate, and the shutdown in the SFR that occured 10 Gyr ago would 
only be modeled by a decrease in the accretion rate and a decrease in the gas density. 
A variation of the star formation efficiency implies that the accretion rate and a direct relation, through the SK law for example,
cannot be the sole or even the most significant factor determining the SFH. Other factors must regulate the star formation rate via 
the star-formation efficiency, such as mechanisms to enhance turbulence like a bar which we invoked to explain the quenching episode of the MW
\citep{haywood2016a}, or stellar feedback \citep[e.g][]{dib2011}, or AGN feedback, which may quench or 
enhance star formation, depending on the distance to the black hole \citep[e.g.][]{robichaud2017}.

\subsection{On the accretion history of our Galaxy}

\subsubsection{Accretion at z$>$2}\label{sec:coldacc}


By the early-00s, modelers had
sufficient computing power to calculate both the dynamical properties
of dark matter over cosmological scales in an hierarchical Universe
and the evolution of gas in galaxies regulated by simple heating and
cooling prescriptions. From their simulations, they found that not all,
perhaps not even most of the accreting gas depending on halo mass, would
be heated to high temperatures when passing through the halo accretion
shock but would instead penetrate down into the halo as filaments of gas
and dark matter  \citep{birnboim2003, keres2005, ocvirk2008, keres2009, agertz2009}.
Within this framework, galaxies with the
halo mass of the MW acquire most of their gas early (z$\ga$2) in their
evolution and largely through the cold mode of accretion \citep{dekel2006,
woods2014, tillson2015}, a fundamental requirement to form a massive thick disk of age greater than 10 Gyr.

Describing an 'instantaneous accretion', the CBM differs from this picture and if the SFR was directly dependent on gas inflow, 
continuous accretion to z$\sim$2 will produce a model different from the CBM. As mentioned above, if the deficit in the number of stars
seen at [Fe/H]<$-0.5$ dex compared to the CBM is confirmed, it may correspond to a 'G-dwarf' problem moved to z$>$3, possibly testifying of a mismatch 
between the CBM and substantially longer-lasting gas accretion. At metallicities above [Fe/H]$\sim$-0.5 dex, the model is a satisfactory match to the data, and it is interesting to try to understand why. The first possible answer is that accretion was fast enough (essentially terminated within z>3) that the CBM is a good approximation starting from z$\sim$3. The second possible answer 
is that the building of the disk -- its star formation rate -- was not directly dependent on gas accretion, as we now speculate.

Several studies have also shown that in the phase where the cosmic
star-formation rate density increases, z$\ga$2, the volumetric gas
accretion rate exceeds the star-formation rate of the ensemble of
galaxies, resulting in a general ``gas accumulation phase'' \citep{dave2012}
or ``gas accretion epoch'' \citep{papovich2011}. During this epoch, the
reservoir of gas is filled up due to the difference in the accretion and
star-formation rates.  Observation show that MW-mass galaxies at redshift
z$\sim$1.3 have molecular gas fractions of $\sim$40-50\% \citep{tacconi2013,
dessauges2015, saintonge2013, papovich2016}.  This implies MW-mass galaxies have
already acquired all the gas necessary to make present MW-type galaxies
and this is without considering the substantial impact of mass return
\citep{leitner2012} (see section \ref{sec:accrez=1}).

Perhaps more importantly, several studies have argued that once
galaxies approach the peak of their star formation at z$\sim$2-3,
they adopt a quasi steady state balance between star formation, gas
accretion, and outflows \citep{bouche2010, dutton2010, dave2012, krumholz2012,
lilly2013} and the gas mass remains approximately constant \citep{dave2012}.
Numerical simulations seem to show a similar phenomenon \citep{stinson2015}.
What these models express is that at these epochs, the SFR is not simply
proportional to the gas supply \citep{lehnert2015}, and some mechanism
breaks a direct coupling between star formation and the gas accretion
rates \citep[e.g.][]{peirani2012, lehnert2015}.  Feedback is offered as the
mostly likely plausible solution, as it can limit the SFR to values
compatible to the observed one \citep[][their Figure~4]{lehnert2015,
agertz2015} and in fact, the MW likely drove significant outflows in its
early evolution \citep{lehnert2014}. Moreover, outflows and large-scale
gas recycling (fountains) provide explanation of the elemental abundances
of the outer disk and its connection with the metallicity of the inner
disks \citep{haywood2013}.  Simulations by \citet{perret2014} suggest that
stellar feedback and accretion might provide a mechanism to saturate SF
in high redshift galaxies by maintaining a high level of turbulence and
fragmentation in the gas.  
In disks already saturated with gas, the variations of its supply may not be the driving 
factor for the level of the SFR and its variations, as suggested previously.
Hence it is possible that the CBM described here provides a good representation of the data between $2<z<3$
because it captures two fundamental features 
of accretion of gas through cold flows: (1) concentrating huge amount of gas rapidly in the central part of disks
(2) lifting the direct dependance of SFR on gas accretion rate.
Finally, if gas accretion continues to z$\sim$2, our description is valid if its impact on the gas out of which stars are forming in the inner 10 kpc of the Galaxy is limited. At these epochs, according to \cite{vandevoort2011}, cold and hot gas accretion brings similar amount of gas into halos. The accretion onto galaxies themselves is roughly a factor of 3 to 4 lower. The cooling time of the hot halo gas is long and thus the accretion rate is perhaps low (e.g., Maller \& Bullock 2004). The cold accreted gas will have higher angular momentum with decreasing redshift and will be preferentially accreted onto the outer parts of the disk \citep[e.g.,][]{danovich15}.  Such gas may move into the inner disk due to, for example, tidal torques from the inner disk but over a longer timescale than its accretion time scale. Moveover, the rate of cold gas accretion onto galaxies will likely be mitigated by feedback from stars in the galaxy itself (e.g. Dubois et al. 2013; van der Voort 2016). Interestingly, star formation intensity in the MW at z$\ga$2 was strong enough to generate significant outflows (Lehnert et al. 2014). Hence, a complex interplay between inflows and outflows generated by feedback could lead to an equilibrium between gas supply and removal. So while qualitatively there are reasons to \textit{a priori} suspect that gas accretion had a limited impact on the effect of chemical evolution of the inner 10 kpc of the disk, to access this possibility quantitatively will require a more detailed model and simulation.

\subsubsection{The quenching episode and the accretion after z$\sim$2}\label{sec:accrez=2}

Simulations suggest that for MW mass halos, the gas accretion rate
peaks at redshifts around 2-3 and then declines fairly steeply
thereafter \citep[e.g.,][]{vandevoort2011, woods2014}.  It is after
the peak that the MW started to quench its star formation
\citep[Fig.~\ref{fig:modeldist};][]{haywood2016a}. While the decline in
the gas accretion rate coincides with quenching epoch in the evolution of
the MW, representing also the change from thick to thin disk formation,
the decline itself is, strictly speaking, probably not responsible for
either the quenching or the change for thick to thin disk formation. In
our model, there is still plenty of gas in the MW disk during this epoch,
of-order 40-50\% of the total mass as is typically observed on MW-type galaxies at z$\sim$1.5
\citep[e.g.,][]{daddi2010, genzel2015}.  If star formation proceeded with an
efficiency typical of disk galaxies at high redshifts, it would
have proceeded unabated.  So some other process(es) must have  triggered this quenching 
phase which is (are) not directly related to the decline in
the gas accretion rate at z$\la$2. \cite{haywood2016a} proposed that
the bar, by sustaining significant turbulence in the gaseous disk, must
have prevented SF from occurring at a typical efficiency.  \cite{khoperskov2017} show, using N-body hydrodynamic simulations, that
it is possible for the bar to suppress the star formation efficiency.
It is most probable, however, that the birth of the bar was made
possible by the disk settling due to the decrease of the gas fraction
in the disk, passing from turbulent disk to a marginally unstable disk
\citep{hayward2017, ma2017, ceverino2017}.  The decrease in the gas fraction,
is, in turn, probably related to the decrease of the gas accretion
rate. That means that although the end of the cold accretion flows is
not directly responsible for the quenching, the two could be causally, but
only indirectly related. This also would explain why these events, decrease
in the gas accretion rate, the formation of a bar, and quenching, all
seem to occur at approximately same epoch.

The thick and thin disks, although separated in time by about 1.5-2 Gyr,
show continuous chemical properties, and in particular a continuous level
of alpha element abundance (in the sense that the thin disk starts at the
same [$\alpha$/Fe] where the thick disk ends; Fig.~3).  For gas accretion
rate to be in equilibrium with star formation rate of the post-quenched
thin disk requires a gas accretion rate of $\sim$ 3 M$_{\odot}$ yr$^{-1}$
\citep[which is also the approximate rate in simulations,][]{woods2014}. Assuming
that the amount of molecular gas in the disk at the end of the thick disk
formation (z$\sim$1.5) was 40\% of the total baryonic mass of the disk
(which we take as 5.10$^{10}$ M$_{\odot}$) and the metallicity of the ISM
at this epoch was about solar, if the Galaxy continued to accrete gas at
a rate of 3 M$_{\odot}$ yr$^{-1}$ during the $\sim$2 Gyr long quenching
episode, then $\sim$0.6 10$^{10}$ M$_{\odot}$ of gas would have been
added to the ISM.  If the metallicity of this newly accreted gas was
[Fe/H]=$-$1 dex, then the ISM of the Galaxy at the beginning of the thin
disk formation would have been 0.1 dex less metal-rich than the ISM at
the end of the thick disk phase (a gas at a metallicity of $-$2 dex would give
a similar value). If the accreted gas was devoided of alpha elements,
the age-alpha relation of Fig. 3 would show a hiatus or an offset of $+$0.1 dex in
[$\alpha$/Fe] at $\sim$7 Gyr compared to the data. Higher amount of alpha
elements would only increase the offset. A hiatus
of this magnitude is not observed.  The measured dispersion in observed
[$\alpha$/Fe] at ages less than 8 Gyr (excluding the 3 sigmas outliers)
is only $\sim$0.02 dex, and even this is an upper limit because it is
consistent with being entirely due to observational uncertainties. In
practice, with these assumptions, to keep variations in the [$\alpha$/Fe]
ratio below 1-$\sigma$ dispersion, any accretion must have had a rate
$<$0.6 M$_{\odot}$ yr$^{-1}$. This effectively means that gas accretion
in the inner disk had come to an end.

\subsubsection{The case for negligible accretion in the MW inner disk
(R$<$7 kpc) since z=1}\label{sec:accrez=1}

Models showing galaxies that form inside-out have shorter infall
timescales in the inner regions compared to the outer regions, and so
an age distribution weighted towards older stars in their inner disks.
Because models parametrize the SFR by Schmidt-Kennicutt type laws,
the resulting SFHs follow the accretion history which are a steep
function of time and thus produce a negligible amount of young stars
in the inner regions.  In \citet{minchev2014}, for example, the disk at
3-5 kpc is overwhelmingly dominated by old, $>$6 Gyr, stars. Since the
SFH follows the accretion history, these models are unable to produce
sufficient numbers of ``young'' (age$<$8 Gyr) and hence metal-rich
([Fe/H]$>$0 dex), stars to reproduce the peak visible in the data at
[Fe/H]$\sim$+0.25 dex. Similarly, in standard models of the bulge, where
this component is represented by exclusively old populations because the
accretion timescale is assumed to be short, which is also what we assume
(but without assuming a relationship between gas mass and star formation
rate surface densities), induces the very fast conversion of gas to stars.

At larger radii in these types of models the accretion timescale is assumed
to be long, and thus most of the stellar mass is formed after z=1.0. This
is discussed for example, in \citet{fraternali2013}, and it is interesting
to compare their scheme with ours.  \citet{fraternali2013} states that
``most of the star formation and therefore most of cold accretion must
take place during the hot-mode phase'' (or at z$<$1-2), which is clearly
at odds with our findings.  Assuming that the thick disk is part of their
description, only about 25\% of the stellar mass of the disk is formed
at z=1.0, which is probably incompatible with our SFH and the massive
thick disk we find. In addition, while they describe the progressive
shift of the accretion towards the outer disk, which we also expect, in their scheme, 
in the inner parts (R$<$6 kpc) most of the gas is accreted at times $<$8
Gyr, and only in the last 2 Gyr the accretion in the outer parts starts
to dominate.  In contrast, we find that to describe the abundances of
the inner disk and bulge, prolonged infall is not necessary. Hence,
we do not find the contradiction discussed by \cite{fraternali2013}
between a cold gas accretion which supposedly dominated the accretion
history at z$\ga$1-2 and a significant fraction of the stellar mass that
would have formed mainly after z=1 in their picture.

At the end of the quenching phase, our model implies that the gas fraction in the Galaxy was
still high, which, coupled to a low SFR, 2-3\Msun yr$^{-1}$, implies a
SFE significantly lower than at the present time.  For a SFR of 2-3\Msun
yr$^{-1}$, a gas fraction of 30-40\% (or 1.5-2 10$^{10}$\Msun) would
yield a depletion timescale of 5-10 Gyr, which, for a disk 7 Gyr old,
suggest that the mass of gas at the end of the
quenching episode was sufficient to maintain a small but significant
star-formation up to the present time (even if all the gas counted in the model 
may not have been available to form stars, see below).  Hence, we would suggest that
prolonged substantial gas accretion is not necessary to explain the
properties of the stellar disk found inside the solar orbit, which
represents most of the mass of the MW. These results also suggest
that the present lack of observed evidence for substantial accretion,
being a factor of 10 below the actual SFR of the MW, is expected.

We emphasize that our description refers to the inner disk of our Galaxy,
and it does not preclude that substantial amount of gas may still be
accreted in the outer parts of the MW in a way described by
gas accretion models at late times. 
Late time accretion would have a relative high angular momentum and could 
result in the extended HI disks observed in the MW and other MW-like galaxies 
\citep[as discussed for example in][]{lehnert2015}.  
Other mechanisms such as a gas accretion drived by energy injection into the halo 
by disk star formation (``galactic fountain'' models) might also be viable mechanisms 
for sustaining disk star formation \citep{armillotta16}.

\begin{figure}
\includegraphics[trim=40 100 40 120,clip,width=10.cm]{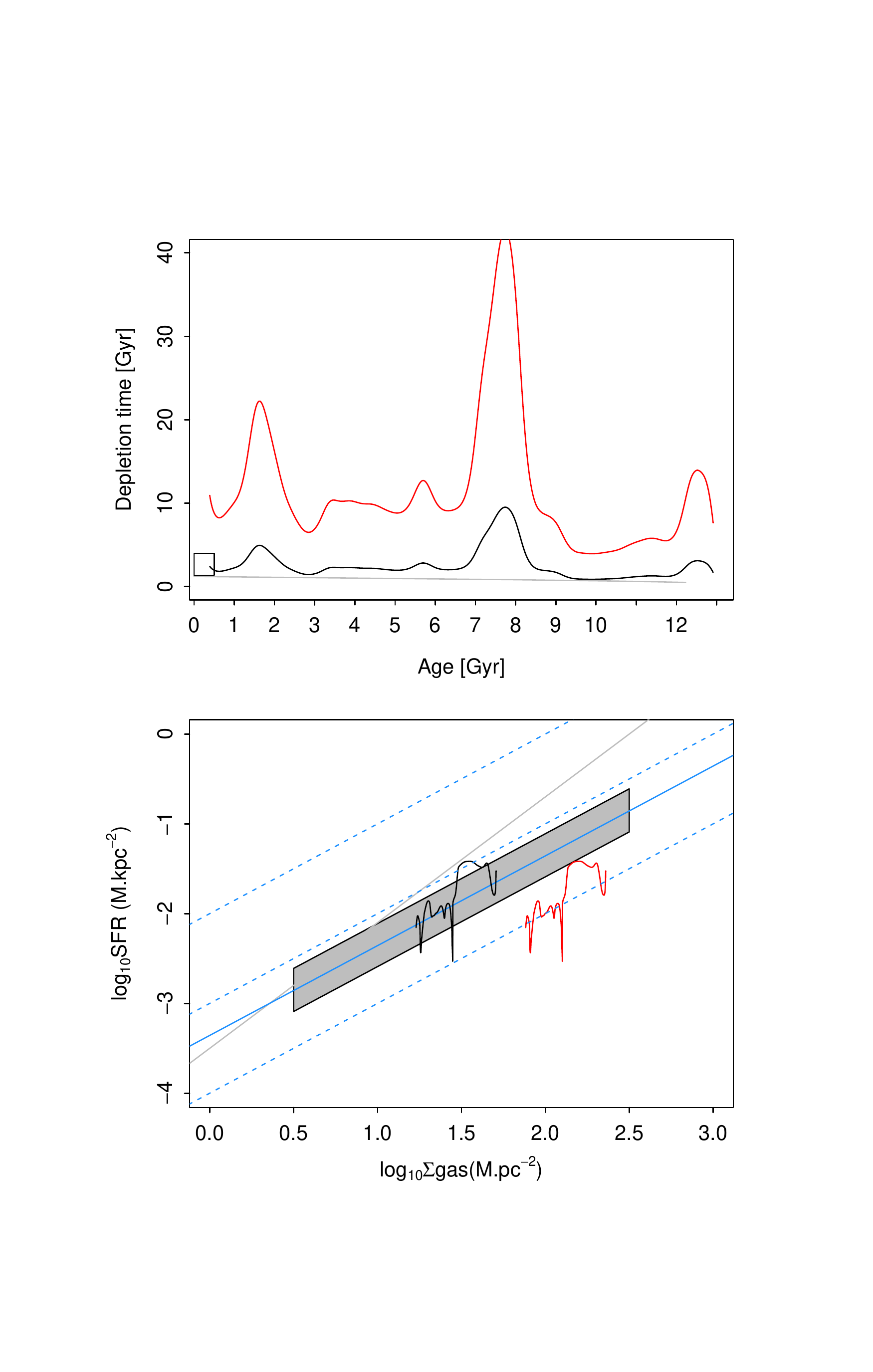}
\caption{Top: depletion timescale of the model, assuming all the gas in the system can form stars, red curve.
The black curve is the depletion timescale of the model, rescaled to have a depletion timescale at present times
compatible with the value of Bigiel et al. (2011) of 2.35 Gyr (square at age=0). It implies 
that at any given time,  2/9 of the gas is molecular. 
The peak at 7.5 Gyr corresponds to the quenching episode.
The gray line is the depletion timescale of molecular gas as a function of time from Tacconi et al. (2017). 
Bottom: Our model on the 
Schmidt-Kennicutt diagram, for the two cases as above. The blue line corresponds to 
the relation of Bigel et al. (2011), and corresponds to a SK law with exponent n=1.
The gray area corresponds to the 1-sigma observations of Bigiel et al. (2011) for H$_2$. 
The gray line is the Kennicutt law with an exponent 1.4.}
\label{fig:sklaw}
\end{figure}


\subsection{Placing our model on the Schmidt-Kennicutt law}\label{sec:sklaw}

We now study how our model is related to the Schmidt-Kennicutt (SK)
relation, keeping in mind that we have a measure of the relative SFH
(Fig.~1). We deduce\footnote{As noted in this paper, our SFH describes the mass growth of the thick disk 
and inner thin disk. Hence, when counting the total stellar mass, we are assuming that
the thin outer disk had a negligible contribution to this mass growth and to the SFH
as discuss in this section. Assuming a thin disk with a scale length of 3kpc,
a limit to the inner thin disk at 7kpc, a very rough estimate gives that we are missing
10\% of the thin disk.} the absolute SFR by fixing the total current stellar
mass of the Galaxy at 5.10$^{10}$ \Msun.  
Our model only considers gas
lost during the evolution of the stellar populations in the MW
\citep{snaith2015}.  This implies that after $\sim$14 Gyr of evolution, 28\%
of the initial total mass of the system is in gas.  To be consistent
with the final mass of the MW implies that the initial baryonic mass
(which is the pool over which the nucleosynthetic products ejected from stars
are recyled) would be 6.4.10$^{10}$ \Msun.

A baryonic census of the MW is difficult.  We have to consider
all the phases.  However, for a CBM to be appropriate,
it must at least be plausible with what is currently known about the
distribution of baryons in the MW itself.  Although very uncertain, the
distribution of mass within the MW is roughly 2.5.10$^{9}$ \Msun in H$_2$,
8.10$^{9}$ \Msun in HI, 2.10$^{9}$~\Msun in HII \citep[][and references
therein]{kalberla2009}. If we naively use these numbers to estimate a gas
fraction of the MW, we find a total mass of 6.2.10$^{10}$ \Msun. There
is of course a hot extended halo whose mass is not well determined. It
could range from insignificant to the dominant gaseous component in the
MW \citep[0.2-12$\times$10$^{9}$~\Msun,][]{miller2013}.  In any event, we
conclude that our CBM is consistent with what is currently
known about the gas content of the MW.

We cannot directly equate the gas in the system that could form stars
using the SK relation. It is already known that for closed-box models,
the application of a SK law gives unrealistic age distribution that peaks
very early (e.g., Fraternali 2012) because of the huge gas surface density
at early times. This is the reason why we did not use the SK law in the
first place to model the star-formation history of the MW.  
At early times, our estimated SFR is $\sim$12-15\Msun.yr$^{-1}$. If SK-type law
were to hold, it implies that either not much of the gas is able to form
stars at any given time or that the star formation efficiency (SFE=SFR/gas
mass) of the MW at those early times was very low.  The depletion time,
t$_{dep}$=1/SFE, of our model, assuming all gas can form stars, is about 4-5 Gyr at early times older than
about 9 Gyr, then rises to $\sim$ 40 Gyr during the quenching episode,
and then is of-order 10 Gyr until today (Fig.~\ref{fig:sklaw}). 
Observations of high redshift (z$>$1) galaxies suggest that the gas depletion time
scales are $\la$1~Gyr \citep{tacconi2013, genzel2015,schinnerer2016,
tacconi2017}. Locally gas depletion timescales are much longer, about
1-3 Gyr \citep{leroy2008, leroy2013, bigiel2014}. These gas depletion
timescales are defined relative to the H$_2$ content of the galaxies
and not the total gas content as we have done. 

In our model, we also assume that not all the gas that makes the reservoir 
in which metals are diluted is molecular, and accept that other modes of the gas, 
such as the hot halo, or HII regions, also take part in efficient mixing.
Hence, we need to rescale our estimates of the gas depletion timescale 
to account for this difference to reconcile them with the
observations of the galaxy population. This can be done in two ways. 
First using
the final gas depletion time of our model and scale it to an estimate of
the gas depletion time of local star forming galaxies.  If we scale our
average gas depletion time during the thin disk formation phase to 2.35
Gyr \citep{bigiel2011}, we must decrease the depletion time history of our
model by about a factor of 4-5 (Fig.~\ref{fig:sklaw}). 
Second, by taking into account only the fraction of gas that participate to 
the star formation, or the H$_2$ relative to the HI and HII masses, which is $\sim$4 in our Galaxy 
according to the above estimates.  
We do not know how this
fraction of the total gas evolved with time, because of the difficulty in measuring
HI masses at high redshifts. First assessments show that within z$\sim$0.2,
this fraction is similar to what is observed on local galaxies
\citep{cortese2017}. At higher redshifts, using absorptions lines as a proxy for estimating the HI content of galaxies suggest an increase with increasing redshifts \citep{noterdaeme2012}. Cosmological
simulations show that this fraction should increase with redshift
\citep[see, e.g.,][]{obreschkow2009,lagos2011,popping2014, dave2017} and also
show that the HI and H$_2$ content may grow in approximate lock-step \citep{dave2017}.
At high redshifts, z$\ga$2, during the thick disk formation phase, using this scaling, the
estimated gas depletion timescale is $\sim$1.0 Gyr, or just below. This is in reasonable
agreement with gas depletion time estimates over those redshifts, see the estimates 
from \cite{tacconi2017} on Fig. \ref{fig:sklaw}.

We can also use the total SFR and gas masses to determine the evolution
of the MW in the gas and star formation rate surface density plane,
the Schmidt-Kennicutt relation. By scaling the SFR and gas mass over the
size of the inner disk during the formation of the thin disk as we did
for the gas depletion time scaling, we can estimate the surface densities
of both.  Once this zero point is set, we can make a relationship between
the surface densities which just then follows the evolution of the SFR
and gas mass (Fig.~\ref{fig:sklaw}). What we find is that the MW evolves
along the local SK relation \citep{bigiel2014} if we apply the same gas
fraction scaling as we did for the gas depletion timescale.  The MW
in its quenching phase, the lowest point on the model curve, appears
as offset from the mean relation, at almost 3$\sigma$, but within the
observed dispersion \citep[][their Fig.~1]{bigiel2014}.

The mode of star formation does not seem to cause significant deviation
from the SK or gas depletion times as long as approximately 1/4 of the
gas is available for forming stars.  At early times, the gas is highly
turbulent and the intense star formation observed in distant galaxies
appears to be marginally unstable \citep[e.g.,][]{lehnert2013}.  After the
quenching episode, the disk settled, and simply grew as a normal spiral
does with a gas depletion timescale of a few Gyr.  During all phases
of evolution, star formation in the MW was inefficient \citep[see][for
a discussion of this point]{papovich2016}. MW-mass galaxies
represent the evolution of $\sim$M$_{\star}$ galaxies over a wide range
of redshifts \citep{ilbert2013}. In fact, MW-like spiral galaxies in
the local universe appear to have very similar star-formation histories
\citep{GD2017}.

We conclude that the Schmidt-Kennicutt relation is an adequate description
of the star formation in the history of the MW as long as most of
the gas at any one time does not participate in the star formation.
Because of the similarities of the MW with other spirals of the same
mass and because MW mass galaxies are approximately M$_{\star}$ over a
wide range of redshifts, this also must be true for the population
of galaxies where a plurality of the stars lie.

\section{Conclusions}\label{sec:conclusions}

We have shown that the bulge and the inner disk (the disk inside the OLR) can be described
by the same chemical evolution, using a closed-box model. 
The model has been compared and is compatible with the MDF, [$\alpha$/Fe]-DF, [$\alpha$/H]-DF, age-metallicity relations, 
as sampled from the APOGEE catalogue from R$\sim$ 6~kpc to the Galactic bulge, microlensing data and inner disk stars sampled at the solar vicinity.
We summarize our main results:\\

\begin{itemize}
\item When identifying the bulge with the disk, only the disk inside the OLR (which we call the inner disk) should be taken into account, because it is only inside the OLR that stars can loose sufficient angular momentum to get trapped in the bar \citep{dimatteo2014, halle2015}. 
Hence, stars at or beyond the OLR are not expected to participate in significant number to building the inner regions \citep{halle2015}. It is therefore expected that the MDF of the bulge and that of the solar vicinity, which appears to be in the OLR region, are different, contrarily to what is sometimes assumed (e.g McWilliam 2016).
\item The chemical properties of the bulge and inner disk -- MDF, age-metallicity relation, chemical patterns of alpha abundances -- 
are well described by the same model, which, combined with dynamical arguments, implies that the bulge is dominated by the thick disk and the thin disk inside the OLR. There is no need for an independent chemical evolution requiring particular parameters to describe most of the bulge.
The model shows an excess of stars at [Fe/H]$<$-0.5 dex, corresponding to ages greater than 11-12 Gyr, when compared to the APOGEE inner disk and bulge MDF. We suggest that this possibly gives the limit below which the CBM approximation becomes invalid.
\item The bimodality observed in the MDF of both the inner disk (after residual contamination by the outer disk stars is removed) and bulge, and the dip in between them stems from the quenching episode that occured in the MW about 9 Gyr ago \citep{haywood2016a} and which separates the thin and thick disks. The dip in the MDF at [Fe/H]$\sim$0 dex is the signpost for a common evolution of the bulge and the inner disk.
\item We infer that the age-metallicity relation in the inner disk and bulge must be much tighter than what is measured in the solar vicinity. The bimodality, or, more exactly, the lull in the MDF at [Fe/H]=0 dex is evidence that the AMR in the inner disk must be tight, otherwise, if the dispersion in metallicity at a given age was significant, we would not see the lull. 
\item Using our SFH, which is determined by fitting the abundance trends with age, allows us not to assume an accretion law, and to leesen the strict coupling which is commonly assumed between the accretion law and the SFR through some Schmidt-Kennicutt star formation 
prescription. What's more, this approach allows us to infer a sudden decrease  of the star formation efficiency which must have occured during the quenching phase (z=1-2), allowing the inner disk to have sustained the same level of SF until the present time without significant replenishment. 
As we argued in \cite{haywood2016a} and further demonstrate in Khoperskov et al. (2017), the formation of the bar permits the decrease of the star formation efficiency, hence of the quenching episode. 
\item The chemical continuity between the thick disk and thin inner disk indicates that the accretion of cold gas cannot have lasted long after the epoch of the beginning of quenching. The continuity in the chemical abundances, and in particular the [$\alpha$/Fe] ratio at the time of quenching imposes strong upper limits on the possible accretion at this epoch and tells us that accretion in the inner disk must have ended at this epoch.
\item The occurence of these three events, the decrease of the accretion rate, growth of the bar, and the quenching episode at approximately the same time is not a coincidence: they are causaly related. 
It is the quenching of the MW star formation activity which allowed the thick phase formation of the Galaxy to end and the thin disk, or secular phase, to begin. The MW gives clear evidence of an example where the morphological transformation is associated to the quenching phase. 
\item  The constraints that we obtain give clues that the gas accretion history of the MW is not the only -- and sometimes not even the main -- parameter determining the variations in the star formation history. 
The establishment of the bar must have induced a significant increase in the depletion timescale (or decrease in the star formation efficiency)  and permitted the inner disk to continue forming stars with no substantial replenishment until the present time. It is therefore possible that during most of its evolution, the inner disk has been overabundant in gas, with the SFR at z$>$2 being limited by feedback and turbulence, and at z$<$1 by a weak star formation efficiency, making the SFH essentially independent of the history of gas accretion.
\end{itemize}

The solar vicinity is just outside the limits of the system described in this paper, being in the region where the OLR is currently located (7-10kpc). We discuss, in a forthcoming paper, how much the evolution of the solar vicinity remains close to the model described here, and quantify the amount of inflow necessary to explain its chemical characteristics.

\begin{acknowledgements}
The Agence Nationale de la Recherche (ANR) is acknowledged for its financial support through the MOD4Gaia project (ANR-15-CE31-0007, P. I.: P. Di Matteo), 
also providing the postdoctoral grant for Sergey Khoperskov.
Francesca Fragkoudi is supported by a postdoctoral grant from the Centre National d’Etudes Spatiales (CNES).
MH and PDM acknowledge the hospitality of Valerie de Lapparent and the 'Galaxies' team at the Institut d'Astrophysique de Paris, where this work has been
partially developed.
\end{acknowledgements}

\begin{appendix}

\section{Abundances}

Abundance ratio of elements have been used as one of the main arguments to argue that the bulge and disk 
are two different populations. The most complete and recent review  is McWilliam (2016), whose
diagnosis in this regard is that the differences are real for several elements. 
In this appendix, we offer a different view and argue that the evidence for significant differences 
are feeble.

\subsection{What level of systematics are expected?}

Table 1 from \cite{mcwilliam2016} shows that the abundance ratio of magnesium at [Fe/H]=0 and [Fe/H]=+0.5 has continuously
decreased  from (respectively) [Mg/Fe]$\sim$+0.28 and [Mg/Fe]$\sim$+0.1 dex in \cite{mcwilliam1994} and \cite{lecureur2007}
to around +0.15 and 0.0 dex in the most recent studies. 
These last values are similar to what is observed on dwarfs of the inner disk observed in the solar vicinity \citep[see e.g.][ Fig. 8, red triangles]{adibekyan2012}.
Systematics between different studies still exist for several elements however. 
The offset measured for instance by \cite{johnson2014} between their measurements and those of \cite{bensby2013} on silicon, calcium, sodium or nickel 
is about 0.1-0.2 dex, while the agreement is good for magnesium and oxygen. 
In addition, while systematic and important effects on metallicities are not expected for solar vicinity
samples, this is still the case for bulge stars. For instance, as mentioned previously, \cite{rojas2014} found that their
metallicity scale is shifted by -0.21 dex compared to the one provided by \cite{hill2011}, which will also
impact the abundance ratios. Hence, differences of the order of 0.1-0.2 dex are not unexpected also on metallicities.

\cite{adibekyan2015} have measured the abundance ratios of different elements for a sample of 
giant stars and shown that these abundances can be offset by as much as 0.1 dex compared to their
sample of solar vicinity dwarfs \citep{adibekyan2012}. This is true for [Si/Fe], but also for
[Mg/Fe] at subsolar metallicities (with similar offsets). Fine-tuning their line-list, \cite{adibekyan2015}
were able to reduce significantly the offset between the two abundances. 
However, the offset they obtained should probably be considered as realistic estimates of the differences to be expected 
when analysing stars that most possibly have the same chemical patterns but are of different stellar types and/or luminosity classes. 
Finally, in a detailed study analysing the spectra of 4 (dwarf) stars by different groups, \cite{hinkel2016} find systematic
differences in temperature and [Fe/H] of more than 100K and 0.2 dex. The differences in [Fe/H] are still of the order of 0.1 dex
when the same line list and the same atmospheric parameters are used.

\begin{figure*}
\includegraphics[width=15.cm]{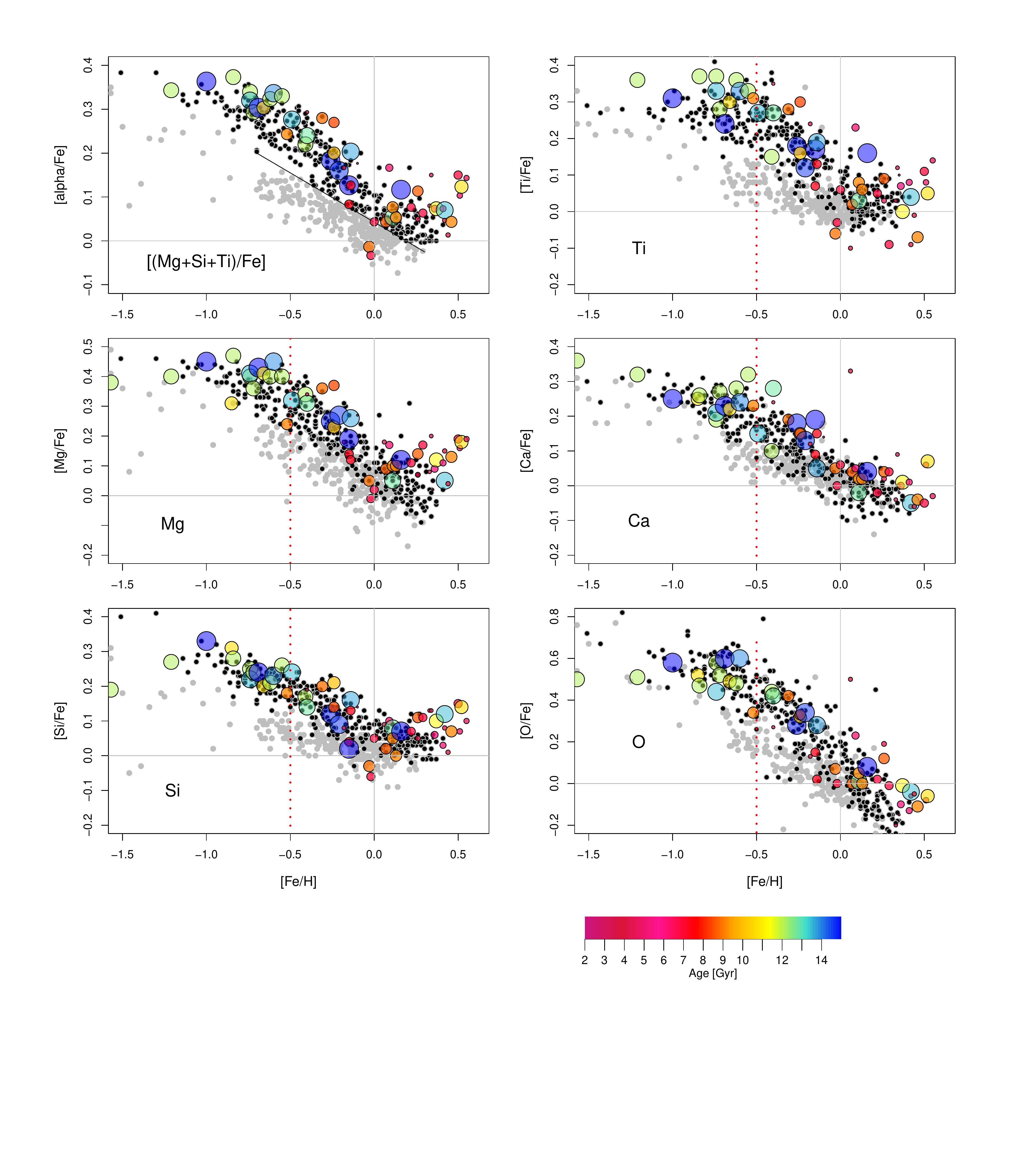}
\caption{Bulge dwarfs and subgiants abundances from Bensby et al. (2013 coloured circles), overlaid
on stars of the solar vicinity from Bensby et al. (2014, gray and black dots) for alpha elements (Mg+Si+Ti),
Al, Ba, Na, Ni. The bulge data for La and Eu comes from Van der Swaelmen et al. (2016).
Stars from Bensby et al. (2014) belonging to the inner disk sequence stars (and which are expected to contribute 
to the bulge) are selected using alpha abundances by choosing those above the black
line in the plot (a). Discarding the outer disk objects (those below the line), the overlap between the 
inner disk stars of the solar vicinity and bulge stars is excellent. 
Colors and sizes of the symbols for bulge stars is coding their ages, which come from Bensby et al. (2013).
}
\label{fig:bulgealpha}
\end{figure*}

\begin{figure}
\includegraphics[width=9.cm]{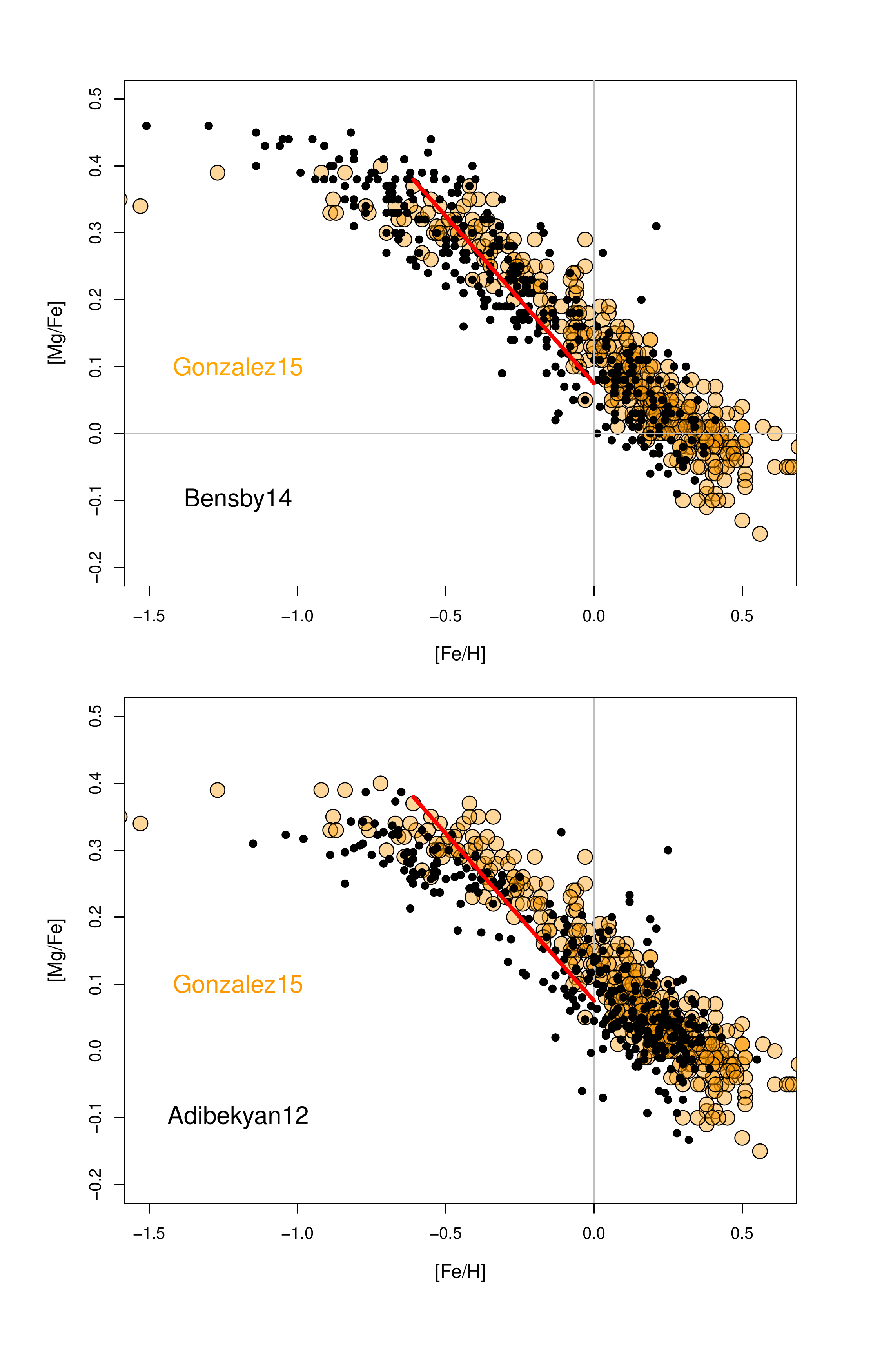}
\caption{Data from Gonzalez et al. (2015) (orange circles) together with ``inner disk'' stars 
of the solar vicinity from Bensby et al. (2014) and Adibekyan et al. (2012), where ``inner disk'' 
stars have been selected in the same way. Note the difference in [$\alpha$/Fe] element sequence 
between Bensby et al. (2014) and Adibekyan et al. (2012). 
See text for comment. 
}
\label{fig:gonzal}
\end{figure}

Below, we examine the arguments that have been developed in the literature as they are summarized and 
updated by McWilliam (2016) in his review, having in mind that differences between studies can still be 
of the order of 0.1-0.2 dex, either in abundance ratios, or in metallicities, either because of residual 
systematics in the analysis, or possible difference between the treatment of dwarfs $vs$ giants.

\subsection{Oxygen}

McWilliam (2016, hereafter McW16) suggests that the knee in [O/Fe] occurs at a higher metallicity in the bulge than in the disk, 
indicating a higher SFR and faster chemical enrichment. His analysis is based on abundances measured by
\cite{johnson2014}, and locates the knee at [Fe/H]=-0.25 dex. The decrease in [O/Fe] occurs at lower
metallicities in Bensby et al. 2013 (hereafter B13) ([Fe/H]$\sim$-0.6 dex), but McW16 seems to favor Johnson et al. because according to him 
alpha elements in B13 also have a knee at higher metallicities. However,
we find no such evidence in Fig. 25 of B13 (see also Bensby et al. 2017), 
where the knee in all elements is certainly at metallicities not higher than [Fe/H]=-0.5 dex
(inasmuch as something like a ``knee'' can really be defined, which, for Si, Ca and even Ti, seems
vain), which means 0.25 dex lower than for oxygen in \cite{johnson2014}. 
This is much larger than any possible difference in the metallicity of the knee of oxygen and 
the other elements in B13.

\subsection{Alpha elements and the case of magnesium}

Fig. \ref{fig:bulgealpha}(a) shows the solar vicinity data for the mean of alpha elements 
Mg, Si and Ti as a function of metallicity, together with the same quantity for microlensed 
dwarfs of Bensby et al. (2014), shown as coloured points. 
On the basis of this plot, we select inner disk stars above the line shown on the plot, with 
OLR and outer stars in grey. 
We then compare the selection of these inner disk stars to the abundances of bulge, microlensed stars 
from \cite{bensby2013} for individual elements: Mg, Si, Ca, and Ti and oxygen on this figure and 
Ni, Zn, Cr, Ba, La, Eu on the next. 

As can be seen from the alpha elements, once the inner disk stars are selected, 
the abundances of the disk and bulge are extremely similar. 
\cite{bensby2013} mentioned that the Mg (and Ti) abundance of the bulge may be slightly 
enhanced compared to his solar vicinity abundances. 
McW16 also advocates that
'for Mg, at least, the bulge trend is measurably different than the thick disk trend.', based on comparison 
between Gonzalez et al. (2015) and \cite{bensby2005} and \cite{reddy2006} for the solar vicinity abundances.

As mentioned above, table 1 of McW16 shows that, while initially found to be systematically
higher than the abundances measured in the solar vicinity, the bulge abundance ratio of [Mg/Fe] has decreased 
systematically over the past 10 years, as testified by the [Mg/Fe] value at solar vicinity, 
measured to be +0.28 dex in \cite{mcwilliam1994} or Lecureur et al. (2007), 
to 0.20 dex in \cite{hill2011}, and 0.15 in \cite{bensby2013}, \cite{johnson2014} or \cite{gonzalez2015}.
Given the enornous range of variations in the measurements of abundances
in the last years, a difference of $\sim$0.05 dex is not solid evidence 
of a different evolution of the bulge and thick disk, also given the fact that McW16 compares bulge abundances
not with the most recent data of the solar vicinity, but with Bensby et al. (2005) and Reddy et al. (2006), 
which have abundance patterns that are much less well defined than in the most recent studies \citep[e.g.][]{adibekyan2012, bensby2014}.

Fig. \ref{fig:gonzal} compares the newest (and largest) solar vicinity abundances samples of Bensby et al. (2014) and 
Adibekyan et al. (2012) (black dots), with inner disk stars selected  as described above, and the data of 
Gonzalez et al. (2015) for the bulge. Two differents remarks can be made from these comparisons.
The first one is that the bulge data of Gonzalez et al. (2015) agree well with those of Bensby et al. (2014):
comparing with the newest data, there is no evidence of a systematic difference between the bulge and the solar 
vicinity.
The second is that significant differences are still visible between the dwarfs abundances of two
different -- but state-of-the-art -- studies of solar vicinity stars, the magnesium abundance ratio of Adibekyan et al. (2012) 
being lower than Bensby et al. (2014) (by $\sim$0.05 dex) at [Fe/H]$<$-0.2 dex. Based on these plots alone, and taking for
granted that differences of 0.05 dex are significant, one would suggest 
that the sample of Gonzalez et al. (2015) and that of Bensby et al. (2014) are the same population (which they may be),
while Bensby et al. (2014) and Adibekyan et al. (2012) are sampling 
two different populations (which they are not).
Hence, if solar vicinity state-of-the-art samples (of dwarfs) still show offsets of about 0.05 dex, why should
we take similar differences between bulge and solar vicinity stars (made, in the case of Johnson et al. (2014), of giants) 
as significant ?

An important reason why McW16 finds the magnesium abundance of the bulge to be enhanced compared 
to the solar vicinity is that he compares with the sample of Bensby et al. (2005), which lacks a clear thick disk 
sequence. Hence, while McW16 finds that [Mg/Fe] at solar metallicity is 0.09 dex higher in the bulge compared
to the solar vicinity stars, we see no significant differences in Fig. \ref{fig:gonzal}.

Some elements have notably suffered systematic offsets, but the situation is improving. 
For instance, Al is measured in Fulbright et al. (2007) to be systematically higher in the bulge 
than in the disk by $\sim$0.2 dex. In \cite{johnson2014} or Bensby et al. (2013), it is well 
compatible with the disk data.

\subsection{Iron peak elements, Zn, Ni}

Fig. \ref{fig:bulgeabund} (left colmun) compares zinc, nickel and chrome from B13 and B14, with the same selection of ``inner disk'' stars
(black dots).

Cu is measured only by Johnson et al. (2012) in the bulge, which Johnson find to differ 
from solar vicinity data. 
Given that other element trends of Johnson et al. (2012) also differ from other bulge data
(this is the case for instance for [Na/Fe], which is measured
to be the lowest at [Fe/H] between -0.8 and -1.0 dex in Johnson et (2012) (at -0.6$<$[Na/Fe]$<$-0.3dex), while it is maximum in Bensby
et al. (2013) with [Na/Fe]$\sim$+0.1 dex at the same metallicities), and the various ways bulge abundances 
have been modified in the recent years, as given in numerous examples in this section, it is safer 
to await for new measurement either to confirm (or not) Johnson et al. (2012) measurements.

\subsection{Neutron capture elements, La, Eu, Ba}

\begin{figure*}
\includegraphics[width=15.cm]{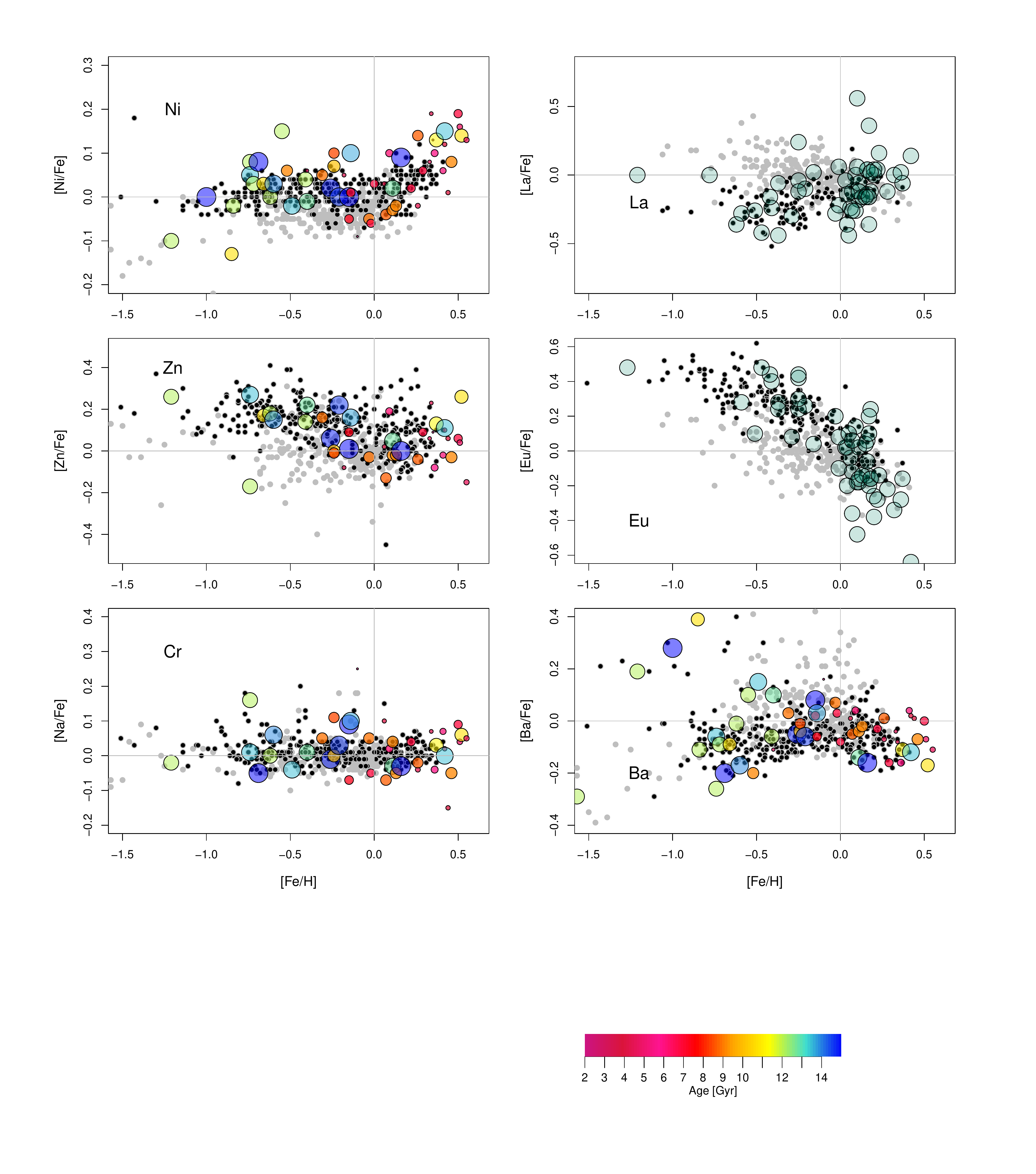}
\caption{
Same as Fig. \ref{fig:bulgealpha} for iron-peak elements (left column) Ni, Zn and Cr, 
and neutron-capture elements (right column) La, Eu, Ba.
The bulge data for Ni, Zn, Cr and Ba come from \cite{bensby2013}, La and Eu from Van der Swaelmen et al. (2016), 
while solar vicinity data for La, Eu come from \cite{battistini2016} and Ba from \cite{bensby2014}.
}
\label{fig:bulgeabund}
\end{figure*}

Fig. \ref{fig:bulgeabund} shows the comparison for Ba, also available from 
Bensby et al. (2013), and Eu and La from van den Swaelmen (2016) for the bulge, with measurements 
of the \cite{bensby2014} sample coming from \cite{battistini2016}. 
Contrarily to \cite{swaelmen2016}, who compare to \cite{bensby2005} and \cite{reddy2006}, 
we do not see that La or Ba are enhanced compared to
thick disk stars, nor that they are underabundant at higher metallicities.

van der Swaelmen et al. (2016) advocated that s-process elements are different 
for bulge and disk stars, the former showing a decreasing trend with increasing metallicity for Ba, La, Ce and Nd, 
being above 0 at metallicities below solar, and below 0 for metallicities above solar. 
The study of Battistini \& Bensby (2016) (not available to van den Swaelmen 2016) shows that these 
trends are well followed by solar vicinity stars, see their Fig. 2, for La, Ce, Nd.
At high metallicities, \cite{swaelmen2016} show lower [Ba/Fe] abundances than B14. 
However, the [Ba/Fe] bulge data of B13 is in perfect agreement with B14, see Fig. \ref{fig:bulgeabund}, 
casting some doubts that the differences between \cite{swaelmen2016} and B14 are real. \\

{(a) [La/Eu]}

Johnson et al. (2012) argued that the ratio of [La/Eu] is lower in the bulge than in the disk and that it 
indicates that the majority of the bulge formed rapidly ($<$1Gyr). This is also claimed by McW16,
comparing the bulge data from Johnson et al. (2012), MFR10 sample, and \cite{swaelmen2016} 
to the more recent solar vicinity measurements of \cite{battistini2016}. 

\begin{figure}
\includegraphics[width=9.cm]{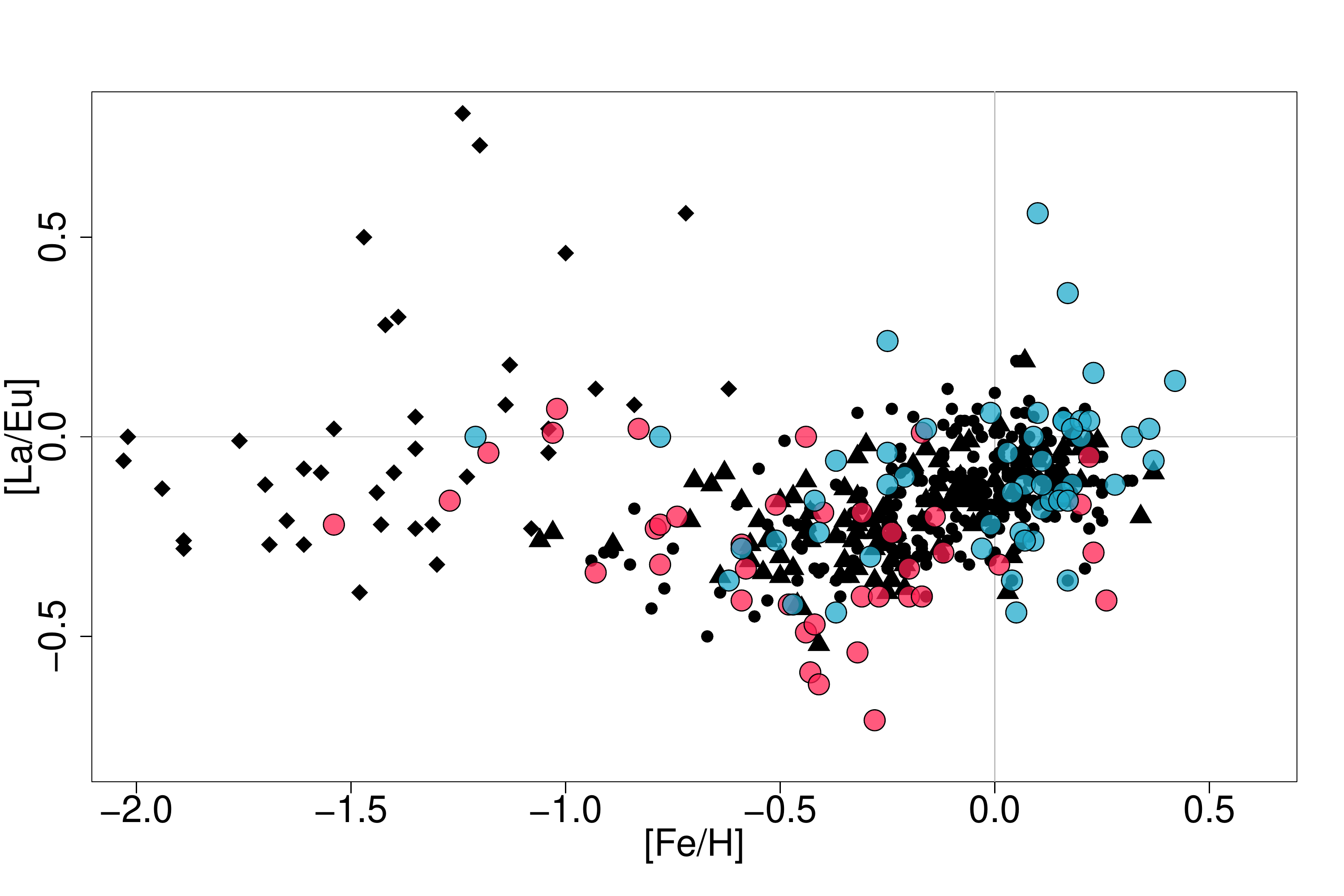}
\caption{[La/Eu] from Johnson et al. (2012) and Van der Swaelmen et al. (2016) (respectively pink and 
blue larger symbols) of bulge stars compared to data from Battistini \& Bensby (2015) (triangles), Mishenina et al. (2013)
(squares) and Ishigaki et al. (2013) (diamons) for solar vicinity stars. 
There is no evidence from these data that the bulge (colored symbols) and the solar vicinity  (black symbols) distributions are different.
}
\label{fig:eula}
\end{figure}

However, we would argue first that the difference between the two bulge samples of Johnson et al. (2012) and \cite{swaelmen2016} 
seem (pink and blue symbols respectively) no less important than the one McW16 sees between solar vicinity stars and Johnson et al. (2012)
(black and pink symbols respectively), and for the 
same reason: Johnson et al. (2012) is clearly undersampling metal-rich stars, which are much better represented in 
\cite{swaelmen2016} or in solar vicinity samples. 
This is illustrated in our Fig. \ref{fig:eula}, which 
compares the  [La/Eu] abundance ratio for the two samples from Johnson et al. 2012 and 
\cite{swaelmen2016}, with three different samples of solar vicinity stars from Battistini \& Bensby (2015),
Mishenina et al. (2013) and Ishigaku et al. (2013). 
There is a good overlap between the bulge (colored symbols) and solar vicinity (black symbols) samples, 
and the two populations are perfectly compatible. 
We conclude that the available [La/Eu] abundance ratios do not provide evidence that the bulge formed 
differently than the inner disk.


\end{appendix}

\end{document}